\documentclass{aa}

\usepackage{graphicx}
\usepackage{txfonts}

\usepackage{hyperref}
\hypersetup{colorlinks=true,allcolors=[rgb]{0,0,0.8}}

\graphicspath{{figures/}{../figures/}}
\usepackage{subcaption}

\usepackage{cleveref}

\crefname{chapter}{Chap.}{Chaps.}
\crefname{section}{Sect.}{Sects.}
\crefname{figure}{Fig.}{Figs.}
\Crefname{chapter}{Chapter}{Chapters}
\Crefname{section}{Section}{Sections}
\Crefname{figure}{Figure}{Figures}

\begin{document}

\title{Low-cost neural summaries for cosmological inference: an ensemble of convolutional neural networks trained on approximate simulations}
\titlerunning{Low-cost neural summaries for cosmological inference}

%
%
%

\author{I. Sáez-Casares\inst{1,2}\corrauth{inigo.saez@unimi.it}        
	\and D. Bianchi\inst{1,3}
	\and M. Calabrese\inst{4}
	\and L. Guzzo\inst{1,3}
}

\institute{
	Dipartimento di Fisica `Aldo Pontremoli', Università degli Studi di Milano, Via Celoria 16, I-20133 Milan, Italy
	\and INFN, Sezione di Milano, Via Celoria 16, I-20133 Milan, Italy
	\and INAF, Osservatorio Astronomico di Brera, via Brera 28, I-20121 Milan, Italy
	\and Astronomical Observatory of the Autonomous Region of Aosta Valley (OAVdA), Fondazione Clément Fillietroz, Loc. Lignan, 39, I-11020 Nus, Italy
}

\date{Received September 30, 20XX}

\abstract
{
	Machine learning can build near-optimal summary statistics for extracting cosmological information from the large-scale structure of the Universe.
	Training them, however, requires large numbers of costly simulations.
}
{
	We aim to build a neural summary trained on fast, approximate low-resolution simulations that remains informative when applied to higher-resolution ones, reducing the cost of the training set.
}
{
	We train an ensemble of convolutional neural networks with different architectures to compress the matter overdensity fields of low-resolution particle-mesh simulations.
	The outputs of a subset of networks are concatenated into a summary statistic, which we use to constrain $\Omega_{\rm m}$ and $\sigma_8$ from high-resolution particle-mesh simulations, each about a thousand times more costly than their low-resolution counterparts.
    Theoretical predictions are provided by a Gaussian process emulator built from the high-resolution simulations.
	We test three strategies for selecting the smallest informative subset of networks.
	Constraints are compared with those from a reference power spectrum analysis. We also consider dark matter halo catalogues from independent high-resolution $N$-body simulations.
}
{
	The neural summary extracts information beyond the binned power spectrum.
	Using multiple networks is essential: a single network can be less informative than the power spectrum.
	A greedy search based on posterior volume provides the tightest constraints with the fewest networks.
	The neural summary also outperforms the power spectrum for dark matter haloes, at different number densities and redshifts, despite being trained on low-resolution matter fields at $z=0$.
}
{
	This work demonstrates that an ensemble of networks trained on approximate simulations yields a summary statistic that remains informative on higher-fidelity data, including dark matter haloes.
	Using multiple networks rather than one is key and offers a route to low-cost neural summaries.
}


\maketitle

\nolinenumbers

\section{Introduction}

The large-scale structure of the Universe is the result of the interplay between gravitational collapse and the expansion of the Universe.
This expansion is accelerated at late times by an unknown dark energy component, which in the $\Lambda$CDM model corresponds to a cosmological constant $\Lambda$.
Current galaxy surveys such as the \textit{Euclid} space mission \citep{Mellier_2025}, the Dark Energy Spectroscopic Instrument \citep[DESI;][]{desi_part_1,desi_part_2}, the Legacy Survey of Space and Time (LSST) at the Vera C. Rubin Observatory \citep{Ivezic_2019}, or the \textit{Nancy Grace Roman} space telescope \citep{Akeson_2019} are already probing, or will soon probe the distribution of galaxies with unprecedented statistical accuracy.
This motivates the design of new robust statistical tools to maximise the amount of information extracted from these data, while minimising the impact of observational and modelling systematics.

To derive cosmological constraints from the spatial distribution of galaxies, it is common to compress the three-dimensional galaxy density field into summary statistics, such as the power spectrum or the two-point correlation function.
Higher-order statistics, such as $N$-point functions ($N>2$), can access non-Gaussian information present on nonlinear scales \citep{Bernardeau_2002} and break degeneracies between galaxy bias and cosmological parameters \citep[see e.g.][]{Veropalumbo_2021}.
Higher-order information can also be accessed through alternative statistics, such as wavelet scattering transforms \citep{Mallat_2012,Cheng_2020}, density splits \citep{Friedrich_2018,Gruen_2018,Paillas_2021}, marked statistics \citep{White_2009,White_2016,Massara_2021}, or augmented correlation functions \citep{Bianchi_2026}, among others.
These statistics typically deliver more compressed and computationally tractable summaries than the hierarchy of $N$-point functions.
However, they are either complicated to model on small nonlinear scales, where most of their advantage with respect to standard summary statistics resides, or completely lack analytical theoretical predictions.
For this reason, their applications to cosmological inference usually require the use of numerical simulations.

Two main options exist for deriving posterior distributions for the cosmological parameters from a given summary statistic using numerical simulations.
One can assume a Gaussian likelihood function, as is commonly done for standard summary statistics.
In such a case, the required ingredients are a numerical covariance matrix computed from simulated mocks and an emulator trained on numerical simulations to capture the cosmological dependence of the summary statistic \citep[see e.g.][]{Cuesta-Lazaro_2024,Valogiannis_2024}.
Alternatively, likelihood-free inference methods do not assume an explicit shape of the likelihood and learn it directly from simulations, relying on methods such as neural density estimation \citep[see e.g.][]{Cranmer_2020,Jeffrey_2021, Hahn_2024, Lanzieri_2025,Jeffrey_2025}.

Machine learning techniques can be used to build near-optimal summary statistics.
These neural summaries are trained to compress the maximum amount of information from density fields into a small number of coefficients.
The compression must be learnt from simulated data, as obtained, for instance, from $N$-body or hydrodynamical simulations.
One of the most common approaches to building neural summaries is to use convolutional neural networks \citep[CNN;][]{LeCun_1989}, which have been used for cosmological inference from simulated density fields \citep[see e.g.][]{Ravanbakhsh_2016,Fluri_2018,Gupta_2018,Ntampaka_2020,Kacprzak_2022,Ntampaka_2022,Villaescusa-Navarro_2022,Min_2024,Sharma_2024,Lanzieri_2025,Cagliari_2026,Saez-Casares_2026,Scelfo_2026}, as well as from real data from weak lensing \citep{Fluri_2019,Jeffrey_2021,Jeffrey_2025} and spectroscopic galaxy clustering \citep{Lemos_2024} surveys.
Other types of neural networks have also been applied to cosmological inference, such as graph neural networks \citep[GNN;][]{Battaglia_2018,Makinen_2022,Villanueva_Domingo_2022,de_Santi_2023,Shao_2023,Balla_2024,Lee_2025}, and, more recently, vision transformers \citep[ViT;][]{Dosovitskiy_2020,Hwang_2023,Gondhalekar_2024,Kakadia_2025}.
Neural networks can also be used to design optimal marks for marked statistics \citep{Semenzato_2026}.

The construction of neural summaries to extract cosmological information from real large-scale structure data faces a computational challenge: generating the required training data is expensive.
Indeed, one requires thousands or tens of thousands of realistic numerical simulations \citep[see e.g.][]{Bairagi_2025}.
Recently, there have been some efforts towards reducing the computational cost of this procedure.
For example, hybrid strategies combine traditional summary statistics, such as the power spectrum or the bispectrum, on large scales with CNNs that extract information from small scales \citep[see e.g.][]{Makinen_2024,Makinen_2025,Bairagi_2026}.
This effectively reduces the complexity of the networks and thus the required number of training samples.
Transfer learning methods have also been used \citep[see e.g.][]{Saoulis_2025,Saoulis_2026}, where neural summaries are pre-trained on a large ensemble of low-fidelity simulations and refined on a small number of high-fidelity ones.
Alternatively, \citet{Pietschke_2026} train a self-supervised ViT on approximate simulations, which produces more informative summaries when applied to high-fidelity simulations than an equivalent supervised network.

In this work, we build a neural summary statistic from CNNs trained on a large number of affordable low-resolution simulations.
This summary is then used to derive cosmological constraints from high-resolution simulations.
More specifically, we build an emulator for the neural summary from a smaller set of high-resolution simulations and use it within a Gaussian likelihood to derive posterior distributions for the cosmological parameters.
Additionally, we devise a strategy based on neural ensembles to maximise the cosmological constraints.
Finally, although the networks are trained exclusively on three-dimensional matter overdensity fields, we also apply the neural summary to catalogues of dark matter haloes.

In practice, we use three sets of simulations of increasing resolution for three different purposes.
First, a large suite of computationally cheap low-resolution simulations is used to train an ensemble of CNNs to compress a matter density field into a compact summary statistic.
Second, a smaller suite of high-resolution simulations, which have not been used during the CNN training, is used to build an emulator that predicts the summary statistic as a function of cosmology and to estimate its covariance matrix.
These ingredients are then used to derive constraints on the cosmological parameters.
The high-resolution simulations are about three orders of magnitude more expensive than the low-resolution ones.
Third, we apply the neural summary to dark matter halo catalogues from independent high-resolution $N$-body simulations, to assess its performance on a more realistic dataset with statistical properties that are further away from those of the training data.
The different simulation sets used throughout this work are summarised in \cref{tab:sims_summary}.

The outline of this paper is as follows.
In \cref{sec:neural_compression} we describe the use of CNNs to compress three-dimensional density fields.
In \cref{sec:neural_summary_stat} we detail the construction of the summary statistic based on neural ensembles.
In \cref{sec:cosmo_inference} we introduce our cosmological inference pipeline.
Finally, \cref{sec:results} presents our main results and \cref{sec:conclusions} summarises our conclusions.

\begin{table*}
	\centering
	\caption{Summary of all simulation sets used in this work.}
	\label{tab:sims_summary}
	\begin{tabular}{lllllllll}
		\hline\hline
		Set               & Code             & $N_{\rm part}$ & $\Delta x \, \left(h^{-1}\mathrm{Mpc}\right)$ & $N_{\rm cosmo}$ & $N_{\rm real}$ & Field type & Runtime (CPUh) & Purpose    \\
		\hline
		Low-res PM        & \texttt{PySCo}   & $128^3$        & $1.28$ (fixed grid)                           & 8192            & 1 (random)     & matter     & $\sim 0.07$    & CNN        \\
		($\Lambda$CDM)    &                  &                &                                               &                 &                &            &                & training   \\
		\hline
		High-res PM       & \texttt{PySCo}   & $512^3$        & $0.32$ (fixed grid)                           & 256             & 1 (fixed)      & matter     & $\sim 70$      & Emulator   \\
		($\Lambda$CDM)    &                  &                &                                               &                 &                &            &                & training   \\
		\hline
		High-res PM       & \texttt{PySCo}   & $512^3$        & $0.32$ (fixed grid)                           & 1               & 1000           & matter     & $\sim 70$      & Data       \\
		(P18)             &                  &                &                                               &                 &                &            &                & covariance \\
		\hline
		\texttt{e-MANTIS} & \texttt{ECOSMOG} & $512^3$        & $0.01$ (AMR)                                  & 80              & 64 (fixed)     & halo       & $\sim 700$     & Emulator   \\
		($w$CDM)          &                  &                &                                               &                 &                &            &                & training   \\
		\hline
		\texttt{e-MANTIS} & \texttt{ECOSMOG} & $512^3$        & $0.01$ (AMR)                                  & 1               & 384            & halo       & $\sim 700$     & Data       \\
		(P18)             &                  &                &                                               &                 &                &            &                & covariance \\
		\hline
	\end{tabular}
	\tablefoot{
		Set is the simulation set, Code the simulation code, $N_{\rm part}$ the number of $N$-body particles, $\Delta x$ the spatial resolution of the gravitational force solver, $N_{\rm cosmo}$ the number of cosmological models, $N_{\rm real}$ the number of realisations per cosmological model, Runtime an estimate of the computing time required to run a single simulation, and purpose the use of the simulations in this work.
		All simulations share the same comoving box side length of $328.125\,h^{-1}\mathrm{Mpc}$.
		The spatial resolution $\Delta x$ corresponds to the finest AMR level for the \texttt{e-MANTIS} simulations.
		We note that the runtime has been computed using different machines and different parallelisation setups.
		It should therefore be interpreted as an order-of-magnitude estimate and not an accurate benchmark.
	}
\end{table*}

\section{Neural compression}
\label{sec:neural_compression}

\subsection{Training samples}
\label{subsec:train_samples}

\begin{table*}
	\caption{
		Ranges of the cosmological parameters covered by the simulation suites considered in this work and fiducial values for the reference cosmological model.
	}
	\label{table:cosmo_params}
	\centering
	\begin{tabular}{c c c c c c c}
		\hline\hline
		Sim. suite / model          & $w$                       & $\Omega_{\rm m}$            & $\sigma_8$                   & $h$                       & $n_{\rm s}$              & $\Omega_{\rm b}$            \\
		\hline
		\texttt{PySCo} $\Lambda$CDM & -                         & $\left[0.1, 0.5\right]$     & $\left[0.6, 1\right]$        & $\left[0.5, 0.9\right]$   & $\left[0.8, 1.2\right]$  & $\left[0.03, 0.07\right]$   \\
		\texttt{e-MANTIS} $w$CDM    & $\left[-1.5, -0.5\right]$ & $\left[0.155, 0.465\right]$ & $\left[0.6083, 1.014\right]$ & $\left[0.55, 0.85\right]$ & $\left[0.72, 1.2\right]$ & $\left[0.037, 0.062\right]$ \\
		\hline
		P18                         & $-1$                      & $0.3071$                    & $0.8224$                     & $0.6803$                  & $0.96641$                & $0.048446$                  \\
	\end{tabular}
	\tablefoot{
		Sim. suite identifies the simulation suite, model the reference cosmology, $w$ is the equation of state of dark energy, $\Omega_{\rm m}$ the matter density parameter, $\sigma_8$ the amplitude of linear matter density fluctuations on the $8\,h^{-1}\mathrm{Mpc}$ scale, $h$ the reduced Hubble parameter, $n_{\rm s}$ the slope of the primordial power spectrum, and $\Omega_{\rm b}$ the baryon density parameter.
	}
\end{table*}

Our aim is to train a CNN to compress the information content of a three-dimensional field into a compact summary statistic made up of a few coefficients ($\mathcal{O}(10)$).
This requires a large number of training samples.
Our strategy is to use low-resolution simulations, which can be generated efficiently in large numbers.
For this we use the particle-mesh (PM) code \texttt{PySCo}~\citep{Breton_2025}.
More specifically, we generate a Sobol sequence \citep{Sobol_1967} with $8192$ points sampling the five-dimensional parameter space of the standard $\Lambda$CDM model: the matter density parameter $\Omega_{\rm m}$, the baryon density parameter $\Omega_{\rm b}$, the reduced Hubble parameter $h$, the slope of the primordial power spectrum $n_{\rm s}$, and $\sigma_{8}$, the root-mean-square amplitude of linear matter fluctuations in spheres of radius $8\,h^{-1}\mathrm{Mpc}$.
The maximum and minimum values for each parameter are given in \cref{table:cosmo_params}.
For each combination of cosmological parameters, we run a PM simulation with a different random seed for the initial conditions, evolving $128^3$ $N$-body particles in a periodic cubic box of comoving side length $328.125\,h^{-1}\mathrm{Mpc}$.
The gravitational force is computed on a fixed mesh of size $256^3$.

From the particle snapshots at $z=0$, we compute the matter overdensity field, which is defined as
\begin{equation}
	\delta(\boldsymbol{x}) = \frac{\rho(\boldsymbol{x}) - \bar{\rho}}{\bar{\rho}},
	\label{eq:overdensity_def}
\end{equation}
where $\boldsymbol{x}$ is the three-dimensional comoving position, $\rho$ is the matter density field, and $\bar{\rho}$ is the mean density of the entire simulation volume.
We compute the density field $\rho$ by assigning the $N$-body particles onto a $64^3$ cubic grid using a piecewise cubic spline (PCS) mass assignment scheme.
This imposes an effective scale cut at the Nyquist frequency of the grid, $k_{\rm Nyq}\simeq 0.6\,h\mathrm{Mpc}^{-1}$.

The simulations are randomly split into three sets: a training set with $7168$ samples, a validation set with $512$ samples, and a test set with $512$ samples.
The training set is used to optimise the loss function, while the validation set is used for early stopping and hyperparameter tuning.
The test set serves to obtain the final performance metrics for each trained network.

\subsection{Loss function}
\label{subsec:loss_func}

The CNN is trained to compress the gridded three-dimensional density field $\boldsymbol{d}$ into a compact summary statistic $\boldsymbol{t}$ by minimising a loss function.
The compressed summary statistic $\boldsymbol{t}$ is therefore implicitly defined by the choice of the loss function.
The simplest option is to compress the field information into the cosmological parameters themselves, that is, to train the CNN to predict the cosmological parameters of the input field.
This compression has been shown to be sub-optimal, and more sophisticated approaches can extract a larger amount of information from the field \citep[see e.g.][]{Charnock_2018, Sharma_2024, Lanzieri_2025}.
In this paper we focus on the simpler cosmological parameter compression, leaving the exploration of more complex methods for future work.

To use the cosmological parameters as targets for the CNN compression, we use a Gaussian negative log-likelihood (GNLL) loss function \citep{Nix_1994}, defined as
\begin{equation}
	\ell_{\rm GNLL} = \frac{1}{N_{\rm batch}}\sum_{i=1}^{N_{\rm batch}}\frac{1}{N_{\rm params}}\sum_{j=1}^{N_{\rm params}} \frac{1}{2} \left[\frac{\left(t_j^i - \theta_j^i\right)^2}{\mathrm{var}_j^i} + \ln\mathrm{var}_j^i\right],
\end{equation}
where $N_{\rm batch}$ is the number of training samples in a single batch, $N_{\rm params}$ is the number of target cosmological parameters, $\theta_j^i$ is the $j$-th cosmological parameter for sample $i$, $t_j^i$ is the $j$-th CNN output for sample $i$, and $\mathrm{var}_j^i$ is the variance of the $j$-th CNN output for sample $i$ as predicted by the CNN itself.
The network outputs two numbers for each target cosmological parameter, the mean and the variance.
This loss function weights each cosmological parameter by its predicted uncertainty, giving more weight to the parameters that the CNN measures more accurately, which stabilises the training procedure.
This approach has been used in previous cosmological analyses based on CNNs \citep[see e.g.][]{Fluri_2018,Fluri_2019,Kacprzak_2022}, and similar results can be obtained with the choice of other loss functions \citep[see e.g.][]{Villaescusa-Navarro_2022}.
After training, the predicted variance is discarded and only the mean is used in the compressed summary statistic, the sole purpose of the variance being to improve the training procedure.

\subsection{Network architecture}
\label{subsec:net_arch}

We use a CNN architecture similar to that of \citet{Saez-Casares_2026}.
The feature extraction part of the network is made of $N_{\rm conv\_block}$ convolutional blocks.
Each block contains $N_{\rm conv\_per\_block}$ convolutional layers with a $3^3$ kernel, which preserve the spatial dimensions of its input.
Every convolutional layer is followed by a batch normalisation layer \citep{ioffe2015batch} and a rectified linear unit (ReLU) activation function.
At the end of each convolutional block we use an average pooling layer with a kernel of size $2^3$ that reduces the size of the field by a factor of two in each dimension.
In addition, the number of output channels in every convolutional layer of a given block is twice that of the previous block.
After the last block, we use a global average pooling layer which returns a single number per channel.

The regression part of the CNN is made of two parallel fully connected networks, one predicting the target cosmological parameters and the other one their variance.
The fully connected networks are composed of two hidden linear layers, each followed by a batch normalisation layer, a ReLU activation layer, and a dropout layer.
A last linear layer returns the final output of each fully connected network, both of which are concatenated to form the output of the CNN.
Each linear layer contains $512$ units.
To ensure positivity, an exponential function is applied to the output of the variance network.

\subsection{Network training}
\label{subsec:net_train}

The networks are trained by minimising the loss function with the \texttt{AdamW} optimiser \citep{Kingma_2014,Loshchilov_2017}.
We use a \texttt{ReduceLROnPlateau} scheduler, which reduces the learning rate by a constant factor after $10$ successive epochs without improvement in the validation loss.
To avoid overfitting the training set, we use an early stopping mechanism that stops the training whenever the validation loss has not improved for $25$ successive epochs.
The maximum number of allowed epochs is set to $1000$, although in practice this number is never reached.
Finally, we save the state of each network at the epoch with the best validation loss.

We use \texttt{optuna} \citep{optuna_2019} to tune the hyperparameters related to the learning procedure, as well as the dropout rate of the fully connected layers (see \cref{subsec:net_arch}).
The other hyperparameters related to the network architecture are fixed as described in \cref{subsec:neural_summary_stat}.
More specifically, we tune the batch size, the initial learning rate, the reduction factor of the \texttt{ReduceLROnPlateau} scheduler, the $\left(\epsilon, \beta_1, \beta_2\right)$ parameters of the \texttt{AdamW} optimiser \citep[for their definitions we refer the reader to][]{Kingma_2014}, as well as the weight decay coefficient.
For each CNN architecture described in \cref{subsec:neural_summary_stat}, we perform $25$ trials using the same sampling, pruning strategy, and similar hyperparameter prior ranges as in \citet{Saez-Casares_2026}.
The target of the optimisation is the validation loss.
Depending on the CNN architecture, a full tuning run takes between five and ten hours on a single NVIDIA A100 or H100 GPU with $80\,$GB of memory.

\section{Neural ensemble summary statistic}
\label{sec:neural_summary_stat}

As described in \cref{sec:neural_compression}, we optimise a CNN to compress the three-dimensional overdensity fields using low-resolution simulations as training data.
However, there is no guarantee that the resulting summary statistic remains informative when applied to higher-resolution simulations.
Our early experiments showed that multiple independently trained CNNs with comparable performance on the low-resolution samples produced very different cosmological constraints when applied to higher-resolution simulations.
In fact, they often performed worse than the power spectrum, which was not the case on the low-resolution samples.
This is because the CNN is applied to data with different statistical properties from those of the training samples.
Therefore, the compression learnt on low-resolution simulations is not necessarily an efficient one when applied to higher-resolution ones.\footnote{We stress that this affects only the constraining power of the summary statistic. Since the emulator and the covariance matrix are built entirely from high-resolution simulations (see \cref{sec:cosmo_inference}), a compression learnt on low-resolution ones can be sub-optimal, but it will not bias the cosmological constraints, provided that the likelihood of the compressed statistic is correctly modelled (see \cref{app:likelihood_validation}).}
To overcome this limitation, we develop a strategy based on \textit{neural ensembles}, which greatly improves the cosmological constraints obtained on high-resolution simulations from the neural compression learnt on low-resolution ones.

\subsection{Building a summary from a neural ensemble}
\label{subsec:neural_summary_stat}

Instead of using a single CNN to build the summary statistic, we can consider several networks, each with different properties, that is, a neural ensemble \citep[see e.g.][for a review]{Dong_2020}.
An ensemble can be generated in different ways, such as using different weight initialisations, different architectures, different training samples, or, more generally, different training strategies.
In this work, we generate an ensemble of networks by varying some of the hyperparameters related to the CNN architecture.
More specifically, we generate a regular grid of hyperparameter values, considering the space formed by the number of convolutional blocks $N_{\rm conv\_block}\in\{1, 2, 3, 4, 5\}$, the number of convolutional layers per block $N_{\rm conv\_per\_block}\in\{1, 2\}$, and the number of output channels in the first block $N_{\rm chs\_first}\in\{5, 10\}$.
We train an independent CNN for each of these hyperparameter combinations, following the procedure described in \cref{subsec:net_train}.
In the end, we obtain an ensemble made up of a total of $20$ CNNs, each with a different architecture.

We have tested several simple ways of combining the outputs of the different CNNs into a compressed summary statistic, such as taking the mean and variance of the ensemble for each CNN output.
Even though this approach can already improve the resulting cosmological constraints with respect to the case of an individual network, it performs worse than a direct concatenation of the CNN outputs.
We therefore build our ensemble summary statistic by concatenating into a single array the outputs of multiple networks.

In practice, each CNN is trained to output five quantities\footnote{After discarding the predicted variances, whose purpose is only to improve the training (see \cref{subsec:loss_func}).}, since they are optimised to compress the input fields into five cosmological parameters of the $\Lambda$CDM model (\cref{subsec:train_samples,subsec:loss_func}).
Therefore, the summary statistic built using the full ensemble of $20$ CNNs would be made of $100$ coefficients.
However, we find that given the small volume of the simulations considered in this work, the CNNs can only reliably measure the parameters $\Omega_{\rm m}$ and $\sigma_8$ (see \cref{app:res_cnn_training}).
For this reason, we keep only the two CNN outputs associated with these parameters and discard the rest.\footnote{We stress that using the cosmological parameters as compression targets is the simplest choice, but not a mandatory one. The interpretation of the CNN outputs in terms of which cosmological parameters are well measured is only relevant for this specific choice of the loss function.}
Therefore, each CNN contributes to the ensemble summary statistic with two elements, giving a total size of $40$ coefficients if the complete ensemble is used.

\subsection{Ensemble pruning strategies}
\label{subsec:ensemble_pruning_strategies}

Ideally, we want to obtain an optimal ensemble summary statistic from the smallest possible number of networks.
A more compressed summary statistic has, in fact, advantages in terms of covariance computation and emulation.
The process of building this sub-ensemble is commonly referred to as \textit{ensemble pruning}.
The selected sub-ensemble should satisfy three requirements when applied to the higher-resolution simulations.
First, each network should provide an informative compression of the field.
Second, the networks should provide complementary information to each other.
Finally, in the case of a finite number of simulations to train an emulator, the selected networks should have a small emulation error, such that the cosmological constraints are not dominated by it.
Finding an optimal solution to this problem is not trivial, in particular if we want to avoid the use of a large number of expensive high-resolution simulations.

We explore three different approaches to select the optimal sub-ensemble of $N_{\rm CNN}$ networks for the neural ensemble summary statistic:
\begin{itemize}
	\item \textbf{Test set MSE ranking:} we rank the different CNNs by their mean-squared error (MSE) computed on the low-resolution test set simulations (see \cref{subsec:train_samples}), defined as
	      \begin{equation}
		      \label{eq:mse}
		      \mathrm{MSE} = \frac{1}{N_{\rm test}}\sum_{i=1}^{N_{\rm test}}\frac{1}{N_{\rm params}} \sum_{j=1}^{N_{\rm params}} \left(t_j^i - \theta_j^i\right)^2,
	      \end{equation}
	      where $N_{\rm test}$ is the number of samples in the test set.
	      We select the $N_{\rm CNN}$ networks with the lowest MSE.
	      This method does not use any information about the high-resolution simulations.
	      In addition, it does not account for the potential complementarity of the different networks, nor the emulation error.
	      However, it is computationally cheap and simple to implement.

	\item \textbf{Individual posterior volume ranking:} for each of the networks of the ensemble in turn, we derive cosmological constraints using the output of that single network as the summary statistic.
	      Then, we rank the networks by the cosmological posterior volume, which is defined as
	      \begin{equation}
		      \mathcal{V}_{\rm p} = \sqrt{\mathrm{det}\boldsymbol{C}_{\boldsymbol{\theta}}},
		      \label{eq:pvol}
	      \end{equation}
	      where $\boldsymbol{C}_{\boldsymbol{\theta}}$ is the covariance matrix computed from the posterior distribution of the cosmological parameters.
	      We select the $N_{\rm CNN}$ networks with the lowest $\mathcal{V}_{\rm p}$ values.
	      This method uses information from the high-resolution simulations and accounts for the performance of each individual network in the final inference scenario, including their emulation error.
	      However, it does not account for the complementarity between networks.

	\item \textbf{Posterior volume greedy search:} we perform a so-called \textit{greedy search} optimisation using the posterior volume as the target metric.
	      More specifically, starting from the network with the smallest individual $\mathcal{V}_{\rm p}$, at each step we add the network that yields the largest reduction in $\mathcal{V}_{\rm p}$, until reaching the target number of networks $N_{\rm CNN}$.
	      This method accounts both for the performance of individual networks on the high-resolution simulations and for the complementarity between them.
	      It is not guaranteed to find the optimal sub-ensemble of $N_{\rm CNN}$ networks, but it requires significantly fewer inference runs than a brute-force search.
\end{itemize}

\section{Cosmological inference}
\label{sec:cosmo_inference}

We build a cosmological inference pipeline based on a Gaussian likelihood, as is common for standard summary statistics and, in many cases, for alternative ones.
Once the CNNs are trained, as described in \cref{sec:neural_compression}, the procedure to derive cosmological constraints is the same as for any other statistic that relies on numerical simulations to obtain theoretical predictions.
The two main ingredients are a covariance matrix computed from a large number of mocks and an emulator providing theoretical predictions for the summary statistic as a function of the cosmological parameters.
In practice, we apply the neural summary described in \cref{sec:neural_summary_stat} to derive cosmological constraints from a set of high-resolution PM simulations, evolving $512^3$ particles with a $1024^3$ mesh in the same cubic volume of side length $328.125\,h^{-1}\mathrm{Mpc}$.
We focus on snapshots at $z=0$, as for the CNN training.

\subsection{Gaussian process emulator}
\label{subsec:gp_emu}

We produce high-resolution PM simulations with the above configuration for the first $256$ cosmological models of the Sobol sequence described in \cref{subsec:train_samples}.
This time, the initial conditions are identical in all simulations.
This first set of high-resolution simulations is used to train an emulator, which predicts the neural summary statistic as a function of the cosmological parameters.
The training data for the emulator is obtained by applying the different CNNs from the ensemble trained on low-resolution simulations to this high-resolution set.
The number of simulations used to train the emulator, $N_{\rm emu}=256$, is within the range of what existing suites of high-resolution $N$-body simulations for cosmological emulators can offer \citep[see e.g.][]{Angulo_2021,Maksimova_2021,Knabenhans_2021,Moran_2023,DeRose_2023,Chen_2025,Sanchez_2026,Tanaka_2026,Yang_2026}.
We study the impact of $N_{\rm emu}$ on the cosmological constraints in \cref{app:emu_nsims}.

We build a Gaussian process emulator \citep[see e.g.][]{Rasmussen_2005} for the neural summary statistic using the \texttt{e-MANTIS}\footnote{\url{https://gitlab.obspm.fr/e-mantis/e-mantis}} code \citep{Saez-Casares_2024a}.
In practice, we build an independent emulator for each element of the summary statistic using a Matérn-$5/2$ kernel.
We have explored the possibility of training a multi-output emulator that accounts for the correlations between the different target elements \citep[see e.g.][]{Bonilla_2007}.
However, in our experiments, this did not improve the accuracy of the final emulator, while significantly increasing the computational cost.

To estimate the accuracy of the emulator, we use a $k$-fold cross-validation procedure.
The training data are divided into $k$ subsets of equal size.
The emulator is trained on $k-1$ sets and evaluated on the remaining one.
This operation is repeated $k$ times, each time leaving a different fold out of the training.
In practice, we use $k=16$, meaning that each fold is made of $16$ samples.
We note that this procedure is only used to estimate the emulator accuracy and the final emulator is built using the whole dataset.
From the $k$-fold cross-validation we compute a covariance matrix describing the emulation accuracy, which takes the form
\begin{equation}
	\boldsymbol{C}^{\rm emu} = \frac{1}{N_{\rm emu}-1}\sum_{i=1}^{N_{\rm emu}}\left(\Delta\boldsymbol{t}_i-\overline{\Delta\boldsymbol{t}}\right)\cdot\left(\Delta\boldsymbol{t}_i-\overline{\Delta\boldsymbol{t}}\right)^\top,
	\label{eq:covariance_emulator}
\end{equation}
where $\Delta\boldsymbol{t}_i = \boldsymbol{t}_i^{\rm emu}-\boldsymbol{t}_i^{\rm sim}$ is the difference between the emulator prediction and the simulation measurement for the validation sample $i$, and the overbar denotes the average over all $N_{\rm emu}$ cross-validation samples.
This emulator covariance ignores the potential cosmological dependence of the emulator accuracy.
It corresponds to a mean emulation error, which might underestimate the emulator accuracy towards the centre of the parameter space and overestimate it towards the edges.

\subsection{Likelihood function}
\label{subsec:lik_func}

We use a Gaussian likelihood, which takes the form
\begin{equation}
	\begin{split}
		\ln \mathcal{L}(\boldsymbol{d}|\boldsymbol{\theta}) = - \frac{1}{2}
		 & \left(\boldsymbol{t}^{\rm data}(\boldsymbol{d})-\boldsymbol{t}^{\rm theo}(\boldsymbol{\theta})\right)^\top \cdot \boldsymbol{C}^{-1} \\
		 & \cdot \left(\boldsymbol{t}^{\rm data}(\boldsymbol{d})-\boldsymbol{t}^{\rm theo}(\boldsymbol{\theta})\right) + \mathrm{const.},
	\end{split}
	\label{eq:likelihood}
\end{equation}
where $\boldsymbol{d}$ is the density field of the data, $\boldsymbol{t}^{\rm data}$ is the neural summary statistic measured from the data, $\boldsymbol{t}^{\rm theo}$ is the theoretical summary statistic predicted by the emulator as a function of the cosmological parameter vector $\boldsymbol{\theta}$, and $\boldsymbol{C}$ is the covariance matrix assumed to be cosmology-independent.
We have omitted the constant normalisation term.

For this likelihood to be valid, we require that the neural summary statistic follow a Gaussian distribution.
We test this in \cref{app:likelihood_validation} for each CNN individually and for their combination.
We find that not all CNNs satisfy the Gaussian likelihood condition individually.
For this reason, we remove from the pool of available CNNs those that fail this test.
In practice, only a single network out of the initial $20$ needs to be removed.

\begin{figure*}
	\centering
	\begin{subfigure}{0.49\linewidth}
		\centering
		\includegraphics[width=\linewidth]{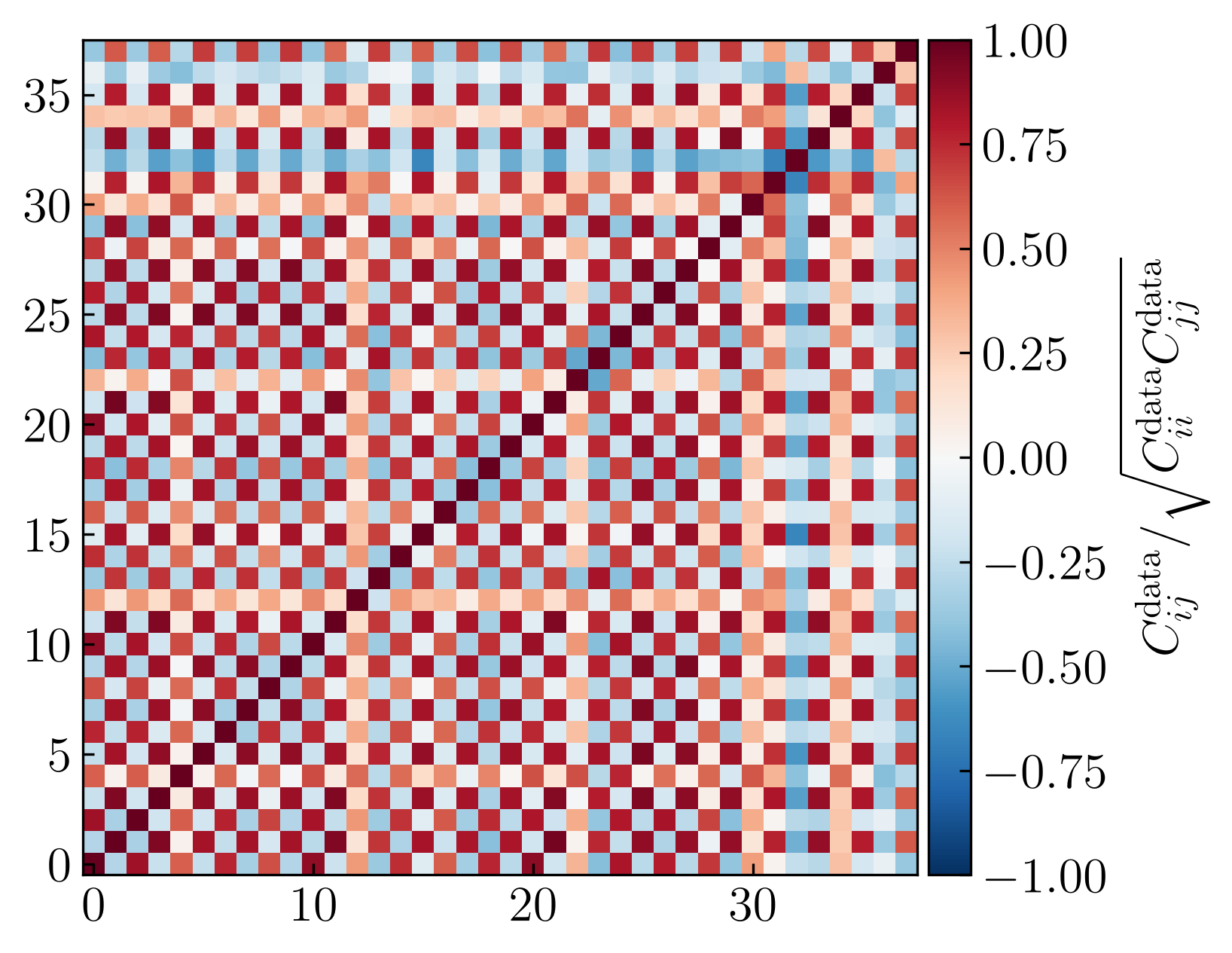}
	\end{subfigure}
	\begin{subfigure}{0.49\linewidth}
		\centering
		\includegraphics[width=\linewidth]{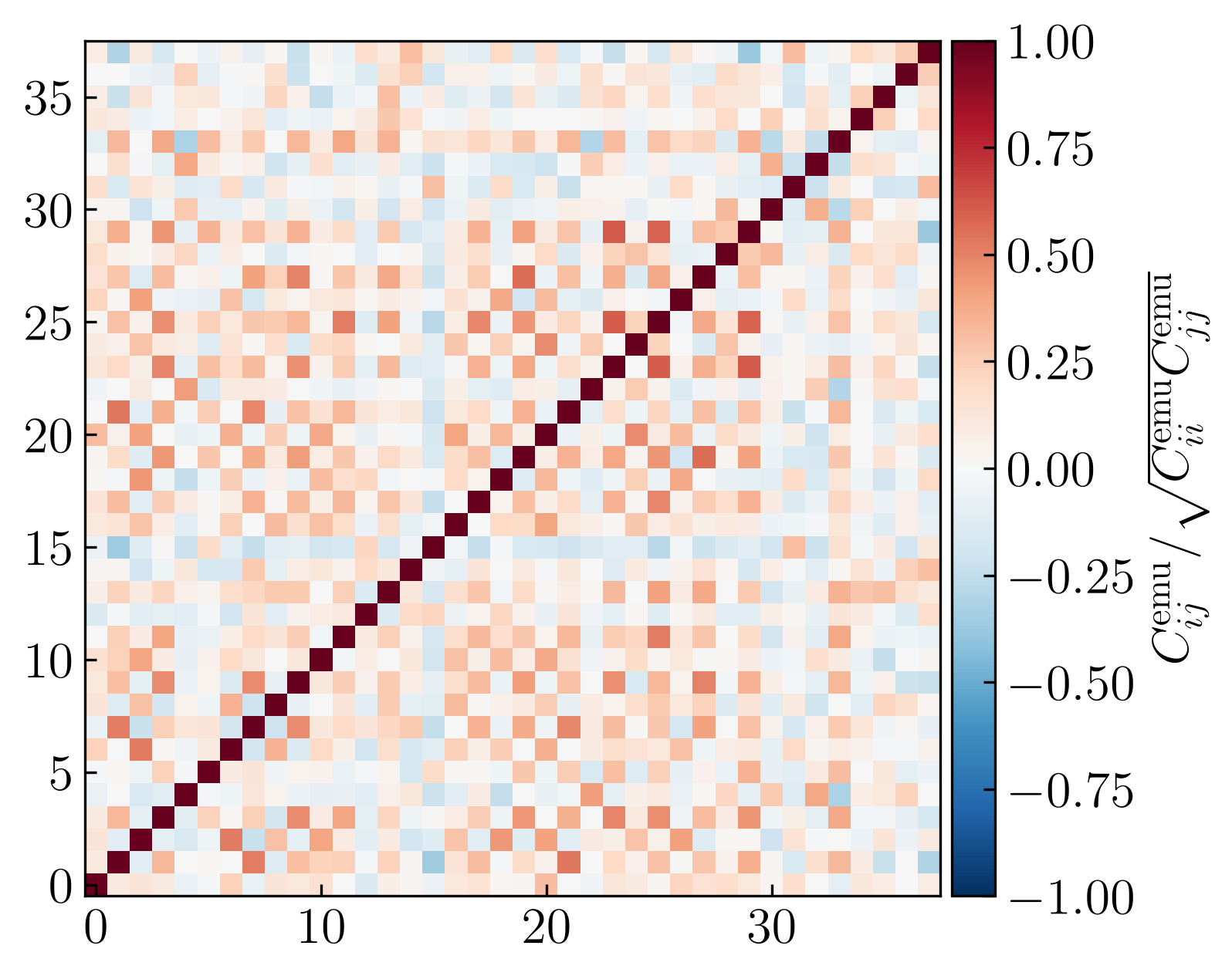}
	\end{subfigure}
	\caption{
		Correlation matrices of the $19$ CNNs that pass the Gaussian likelihood condition when evaluated on the high-resolution PM simulations.
		The CNNs are ordered by their architecture, first varying the number of convolutional blocks, then the number of convolutional layers per block, and finally the number of channels in the first block.
		Left: data correlation matrix.
		Right: emulator correlation matrix.
	}
	\label{fig:covariance_data_emu}
\end{figure*}

The covariance matrix can be written as
\begin{equation}
	\boldsymbol{C} = \boldsymbol{C}^{\rm data} + \boldsymbol{C}^{\rm emu} + \boldsymbol{C}^{\rm sim},
	\label{eq:covariance_total}
\end{equation}
where $\boldsymbol{C}^{\rm data}$ is the covariance associated with the data vector, $\boldsymbol{C}^{\rm emu}$ is the emulator covariance given by \cref{eq:covariance_emulator}, and $\boldsymbol{C}^{\rm sim}$ represents the cosmic variance of the finite-volume simulations used to train the emulator.
To estimate the data covariance, we produce a second set of $1000$ high-resolution PM simulations, distinct from the $256$ used to train the emulator, all sharing a single fiducial cosmology but with different initial conditions.
As fiducial cosmology we use the P18 model from \cref{table:cosmo_params}, which is compatible with the CMB constraints from \citet{Planck_2018}.
In \cref{fig:covariance_data_emu} we show the correlation matrix for the entire ensemble of CNNs that satisfy the Gaussian likelihood condition.
We note that in most cases we will use a subset of this covariance matrix, since in practice we aim to build the optimal summary statistic from a sub-ensemble of the whole CNN set.

The emulator is trained on simulations with the same volume as the fiducial ones.
Therefore, we have $\boldsymbol{C}^{\rm sim}=\boldsymbol{C}^{\rm data}$ in our case.
In a realistic analysis scenario, one would like to reduce the cosmic variance of the emulator training set well below that of the data, such that $\boldsymbol{C}^{\rm sim}$ can be neglected.
In practice, we use as data vector the fiducial realisation whose initial conditions are identical to those of the emulator training simulations, so that the cosmic variance of the data and of the emulator cancel and $\boldsymbol{C}^{\rm sim}$ can be ignored.

\subsection{Cosmological parameter posteriors}
\label{subsec:cosmo_posteriors}

We use the \texttt{Nautilus} sampler \citep{Lange_2023} to obtain posterior distributions for the cosmological parameters using $5000$ live points.
In practice, we only sample $\Omega_{\rm m}$ and $\sigma_8$ and keep the other parameters fixed to their fiducial values.
Indeed, given the small volume of our simulations, these are the only two parameters that can be meaningfully constrained.
We use flat priors delimited by the range of validity of the emulators as given in \cref{table:cosmo_params}.

\subsection{Matter power spectrum reference analysis}
\label{subsec:matter_power_spectrum}

To benchmark the constraining power of our neural summary statistic against standard two-point information, we consider the binned power spectrum as a reference probe.
\texttt{PySCo} computes the matter power spectrum on-the-fly using the same grid as the PM gravity solver.
In all the cases considered in this work, the power spectrum is used up to a maximum wavenumber that is always smaller than half of the Nyquist frequency of the mesh on which it is measured, which mitigates aliasing effects \citep{Sefusatti_2016}.
In practice, we use the power spectrum computed from $k_{\rm min}\simeq0.024\,h\mathrm{Mpc}^{-1}$ up to $k_{\rm max}\simeq0.6\,h\mathrm{Mpc}^{-1}$ with $31$ linearly spaced bins.
The value of $k_{\rm max}$ matches the Nyquist frequency $k_{\rm Nyq}$ of the $64^3$ grid used for the CNNs, such that both probes approximately probe the same scales.

We build an inference pipeline following the same procedure as previously described for the neural summary statistic.
The emulator is built using Gaussian processes, with the only difference that the binned power spectrum is first compressed with a principal component analysis, from which we keep the first $10$ components.
We estimate the emulator covariance matrix with the $k$-fold cross-validation procedure and the data covariance matrix from the same set of $1000$ mocks.
These ingredients are then combined into a Gaussian likelihood.
We note that in most cases the ratio of emulator error to cosmic variance of the data for the power spectrum is comparable to or smaller than that of the neural summary statistic.
This tends to favour the power spectrum probe, which results in a more conservative assessment of the neural summary performance.

\section{Results}
\label{sec:results}

\subsection{Emulator cross-validation}
\label{subsec:res_emu_cv}

\begin{figure}
	\begin{center}
		\includegraphics[width=\linewidth]{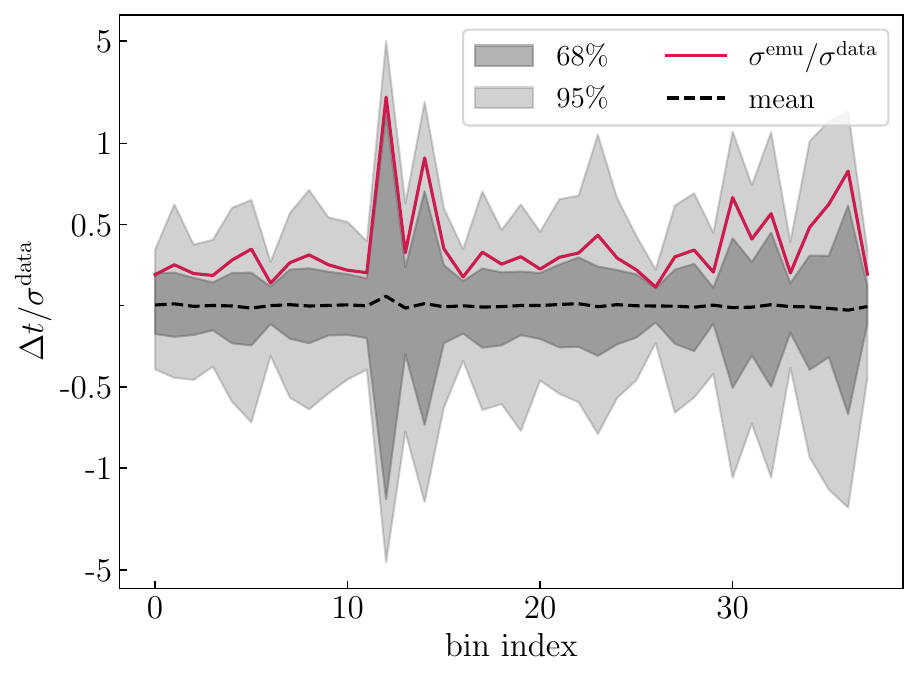}
	\end{center}
	\caption{
		Distribution of the emulation error for each element of the summary statistic considering the $19$ CNNs that pass the Gaussian likelihood condition, normalised by the data standard deviation.
		The $68\%$ and $95\%$ confidence intervals are given by the shaded areas, along with the mean as a dashed line.
		The solid line corresponds to the square root of the diagonal of the emulator covariance used in the likelihood.
		The $y$-axis is linear between $-1$ and $1$ and logarithmic elsewhere.
	}
	\label{fig:emu_err_diag}
\end{figure}

As described in \cref{subsec:gp_emu}, we build an emulator that predicts the neural summary statistic as a function of the cosmological parameters.
We report here the results of the cross-validation procedure that quantifies the accuracy of this emulator.
More specifically, we provide in \cref{fig:emu_err_diag} the distribution of diagonal emulator errors normalised by the data standard deviation.
We consider the ensemble of $19$ CNNs remaining after applying the Gaussian likelihood condition, each contributing two elements to the compressed summary statistic, resulting in $38$ coefficients.
For most coefficients, the $68\%$ confidence interval has a half-width of $\sim25\%$ of the data standard deviation.
We also show in \cref{fig:emu_err_diag} the square root of the diagonal of the emulator covariance, computed from \cref{eq:covariance_emulator}, also normalised by the data standard deviation.
It follows the trend of the $68\%$ confidence level, while being slightly larger for most coefficients.
For some coefficients, the emulator error is close to or larger than the data standard deviation, meaning that the cosmological constraints will be strongly limited by the emulation accuracy in such cases.
Both the individual posterior volume ranking and the greedy search presented in \cref{subsec:ensemble_pruning_strategies} naturally account for this effect when selecting the optimal subset of networks to build the summary statistic.

From the cross-validation procedure we also obtain a full non-diagonal emulator covariance as given by \cref{eq:covariance_emulator}.
We show the corresponding correlation matrix in the right panel of \cref{fig:covariance_data_emu}.
There are non-trivial correlations between the emulator errors of different coefficients.
This is expected given that the data covariance itself has large non-diagonal terms.
However, the amplitude of the non-diagonal terms in the emulator correlation matrix is significantly smaller than that in the data correlation matrix.

\subsection{Ensemble pruning}
\label{subsec:res_ens_pruning}

\begin{figure}
	\centering
	\includegraphics[width=\linewidth]{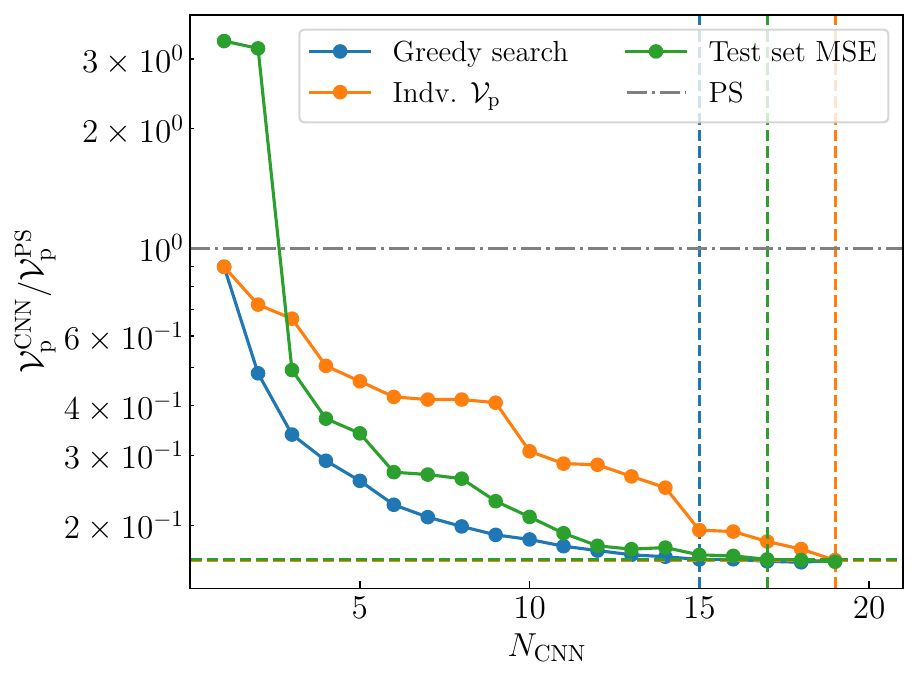}
	\caption{
		Posterior volume ratio with respect to the power spectrum analysis, as a function of the number of CNNs, for different ensemble pruning methods.
		The vertical and horizontal dashed lines mark the convergence point for each method.
		The dash-dotted line gives the power spectrum level, that is, a posterior volume ratio of one.
	}
	\label{fig:pvol_vs_n_cnn}
\end{figure}

We follow the three ensemble pruning methods described in \cref{subsec:ensemble_pruning_strategies}, that is, test set MSE ranking, individual posterior volume ranking, and greedy search.
These different methods give us a way of selecting $N_{\rm CNN}$ networks out of the full ensemble to build the summary statistic.
For each of these methods, we vary the number of selected CNNs from $N_{\rm CNN}=1$ to $N_{\rm CNN}=19$, that is, from a single network to using the full ensemble (only considering those that satisfy the Gaussian likelihood condition).
The objective is to obtain the best possible constraints with the smallest number of networks, so as to work with a more compressed summary statistic.
Indeed, a more compressed statistic has advantages both in terms of covariance matrix computation and emulator training.
We use the posterior volume $\mathcal{V}_{\rm p}$ as defined in \cref{eq:pvol} to quantify the constraining power of the summary statistic, with a smaller $\mathcal{V}_{\rm p}$ corresponding to better constraints.
The posterior volume is computed on the two-dimensional space given by the parameters $\Omega_{\rm m}$ and $\sigma_8$.
The same $\mathcal{V}_{\rm p}$ definition is used as optimisation target for both the posterior volume ranking pruning and the greedy search.
In practice, we consider that the cosmological constraints have converged as a function of the number of networks when the ratio of the posterior volume with respect to the power spectrum analysis $\mathcal{V}_{\rm p} / \mathcal{V}_{\rm p}^{\rm PS}$ is within three per cent of its minimum value.
This threshold is chosen to be larger than the posterior sampling noise, which we have estimated to be of order one per cent.

In \cref{fig:pvol_vs_n_cnn} we show the ratio of the posterior volume $\mathcal{V}_{\rm p}$ obtained with the neural summary statistic with respect to that of the power spectrum, as a function of the number of networks, for the different pruning strategies.
The greedy search approach reaches a plateau at around $N_{\rm CNN}=15$.
The MSE ranking converges more slowly than the greedy search, and the individual $\mathcal{V}_{\rm p}$ ranking does not converge until the full ensemble is used.
Additionally, the greedy search provides the best cosmological constraints at each value of $N_{\rm CNN}$.
This result shows that in order to obtain an informative summary statistic, it is not enough that the individual networks are informative on their own: they must also probe complementary information.
This can be achieved through the greedy search.
We note that the greedy search is not guaranteed to produce the optimal result.
Indeed, a brute force approach testing all possible combinations might find convergence for a lower number of networks.
However, such a method would be computationally expensive.
Therefore, for the rest of this work we adopt the greedy search to build the neural ensemble summary statistic with the smallest number of networks.

It is worth noting that in the case of a single network, the ranking based on the test set MSE does not find the network with the smallest individual posterior volume.
In fact, the first network given by the test set MSE ranking gives a $\mathcal{V}_{\rm p}$ that is $\sim 3$ times larger than that of the power spectrum analysis.
This shows that the individual performance of a network on the low-resolution sample is not representative of its performance when applied to the high-resolution simulations, which, as already discussed in \cref{sec:neural_summary_stat}, is the initial motivation for an ensemble-based statistic.

\begin{figure}
	\centering
	\includegraphics[width=\linewidth]{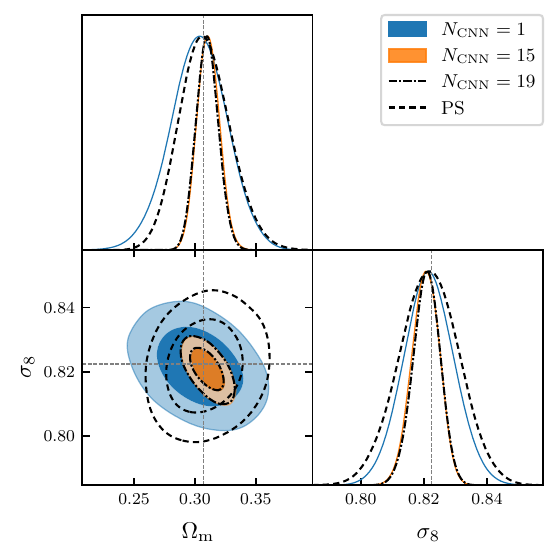}
	\caption{
		Two-dimensional posterior distributions of the cosmological parameters obtained with the neural summary using different numbers of networks, as well as the power spectrum analysis.
		The networks for the neural summary analyses have been selected with the greedy search approach.
	}
	\label{fig:post_n_cnn}
\end{figure}

In \cref{fig:post_n_cnn} we show the posterior distributions of the cosmological parameters for different analysis setups, based on the greedy search approach.
More specifically, we give the posteriors obtained with $N_{\rm CNN}=1$, that is, the individual network with the smallest posterior volume, and with $N_{\rm CNN}=15$, which corresponds to the convergence of the posterior volume as a function of the number of networks.
We also show the case of the complete ensemble $N_{\rm CNN}=19$, which is in good agreement with $N_{\rm CNN}=15$, showing the convergence of the cosmological constraints with the number of networks.
We also show the posteriors from the power spectrum analysis as a reference.
Finally, in \cref{tab:summary} we report the ratio of one-dimensional standard deviations on the cosmological parameters obtained with these different configurations with respect to the power spectrum.
We note that, even though the best individual CNN from the greedy search produces a smaller $\mathcal{V}_{\rm p}$ than the power spectrum, the standard deviation on the individual parameter $\Omega_{\rm m}$ is larger, showing once again the importance of the ensemble combination.
This is also related to the fact that our optimisation metric is the two-dimensional posterior volume, rather than the one-dimensional marginal widths.

\begin{table}[t]
	\centering
	\caption{
		Constraints from the neural summary relative to the power spectrum for different analysis configurations.
	}
	\label{tab:summary}
	\begin{tabular}{lccc}
		\hline
		\hline
		Analysis                     & $\sigma_{\Omega_{\rm m}} / \sigma_{\Omega_{\rm m}}^{P(k)}$ & $\sigma_{\sigma_8} / \sigma_{\sigma_8}^{P(k)}$ & $\mathcal{V}_{\rm p}/\mathcal{V}_{\rm p}^{P(k)}$ \\
		\hline
		CNN: $N_{\rm CNN}=1$         & $1.11$                                                     & $0.85$                                         & $0.90$                                           \\
		\hline
		CNN: $N_{\rm CNN}=19$        & $0.43$                                                     & $0.45$                                         & $0.16$                                           \\
		\hline
		CNN+$P(k)$: $N_{\rm CNN}=19$ & $0.40$                                                     & $0.42$                                         & $0.14$                                           \\
		\hline
	\end{tabular}
	\tablefoot{
		The ratios of one-dimensional standard deviations obtained with different neural summary analyses with respect to the power spectrum alone are given for each cosmological parameter, as well as the two-dimensional posterior volume ratio.
		The case of $N_{\rm CNN}=1$ uses the individual network minimising the posterior volume $\mathcal{V}_{\rm p}$, while in the other cases the full ensemble is used.
	}
\end{table}

\subsection{Combination with the power spectrum}
\label{subsec:ps_comb}

We perform a combined analysis of the neural summary statistic and the power spectrum.
We account for the cross-correlation between the two probes by computing the full data covariance from the set of $1000$ realisations.
For simplicity, the emulator covariance is assumed to be block-diagonal, that is, we neglect the cross-correlation between the emulation error of the two probes.
We have verified that this assumption has a negligible impact on the cosmological posteriors.

\begin{figure}
	\centering
	\includegraphics[width=\linewidth]{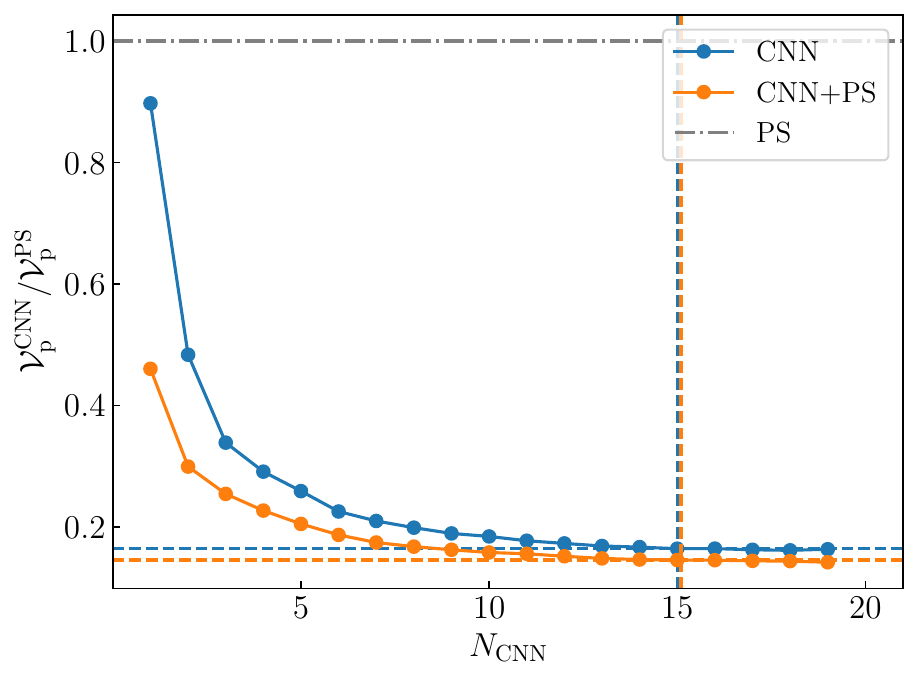}
	\caption{
		Posterior volume for the CNN+$P(k)$ combination and CNN only as a function of the number of CNNs, using the greedy search method.
		The vertical and horizontal dashed lines mark the convergence point for each case.
		The dash-dotted line gives the power spectrum level, that is, a posterior volume ratio of one.
	}
	\label{fig:pvol_vs_n_cnn_ps_comb}
\end{figure}

We repeat the ensemble pruning with the greedy search method, but this time the power spectrum is combined at each step.
This way, by construction, the greedy search selects networks that carry information complementary to the power spectrum.
In \cref{fig:pvol_vs_n_cnn_ps_comb} we show the posterior volume ratio as a function of the number of networks in the neural summary statistic.
We compare it to the case of the CNN-only analysis.
Adding the power spectrum in the greedy search does not lead to faster convergence.
The final posterior volume $\mathcal{V}_{\rm p}$ is $12.5\%$ lower than in the CNN-only case.

\begin{figure}
	\centering
	\includegraphics[width=\linewidth]{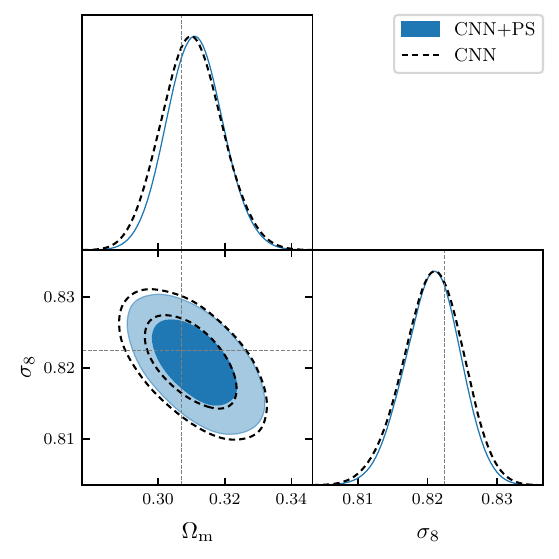}
	\caption{
		Two-dimensional posterior distributions of the cosmological parameters for the combined CNN+$P(k)$ analysis and for the CNN only.
		In both cases, the full ensemble of $19$ networks is used.
	}
	\label{fig:post_n_cnn_ps_comb}
\end{figure}

In \cref{fig:post_n_cnn_ps_comb}, we show the two-dimensional posterior distributions resulting from the CNN+$P(k)$ analysis and the CNN-only one.
In both cases, we use the full ensemble of $19$ CNNs.
The CNN+$P(k)$ case has slightly tighter contours than the CNN-only case.
We also report the improvements in terms of one-dimensional standard deviations on the cosmological parameters in \cref{tab:summary}.
The gain from adding the power spectrum to the neural summary statistic is of $\sim 7\%$ for both $\Omega_{\rm m}$ and $\sigma_8$.
We conclude that the neural summary alone already captures most of the information contained in the power spectrum.

\subsection{Application to dark matter haloes}
\label{subsec:nbody_haloes}

\begin{figure}
	\centering
	\includegraphics[width=\linewidth]{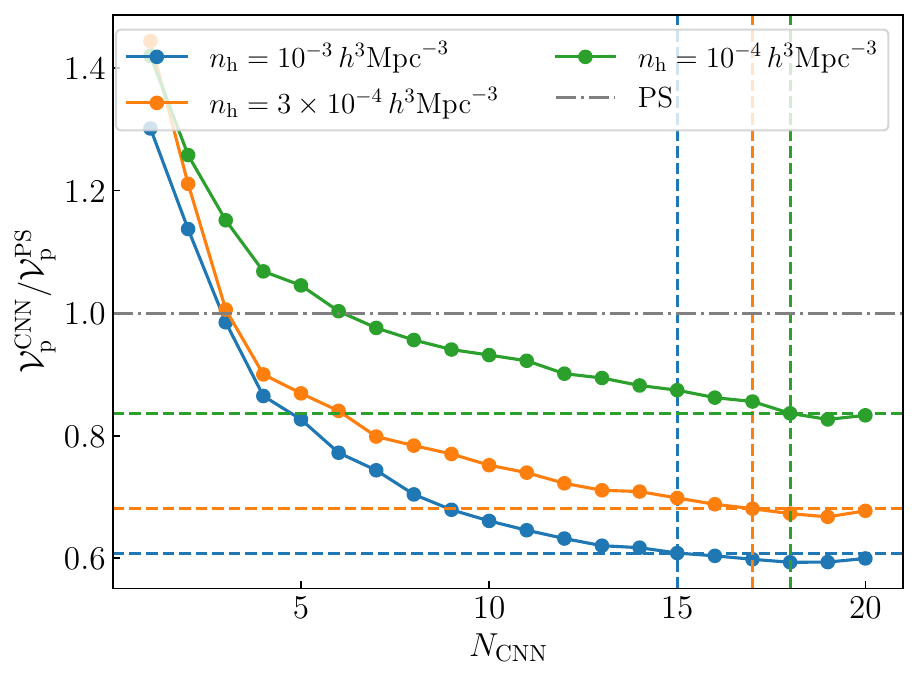}
	\caption{
		Posterior volume ratio as a function of the number of CNNs for different halo number density values at $z=0$, using the greedy search approach.
		The vertical and horizontal dashed lines mark the convergence point for each case.
		The dash-dotted line gives the power spectrum level, that is, a posterior volume ratio of one.
	}
	\label{fig:pvol_vs_n_cnn_emantis_nd}
\end{figure}

We now assess the performance of our neural summary statistic when applied to derive cosmological constraints from biased tracers.
In the context of a real galaxy clustering data analysis, galaxies are usually modelled directly in the simulations used to train the CNNs.
This usually requires identifying dark matter haloes and populating them with galaxies using an empirical prescription, or adopting a galaxy bias model that relates the galaxy overdensity field to the matter one.
In keeping with the aim of this work, which is to minimise the cost of producing the CNN training samples, we study the performance of our neural ensemble summary statistic trained only on matter fields when applied to dark matter haloes.
We note, however, that in the procedure described in this section, the cosmological dependence of the halo bias is captured by the emulator, which is trained on the same halo selection as the data.
For this reason, the halo bias is not a free parameter and the only free parameters are still $\Omega_{\rm m}$ and $\sigma_8$, for both the neural summary and power spectrum analyses.
Therefore, this is not a test of the ability of the neural summary to break degeneracies between halo bias and cosmological parameters, but rather a test of the statistic on a more realistic dataset that is further away from the training set.

In this exercise, the CNNs are unchanged and still trained on the matter fields obtained with low-resolution PM simulations.
We then apply them to overdensity fields built from halo catalogues.
In other words, with respect to the results presented in the previous sections, the high-resolution PM matter fields are now replaced by halo overdensity fields.
The rest of the pipeline remains unchanged.
We use $N$-body halo catalogues from the extended \texttt{e-MANTIS} simulation suite~\citep{Saez-Casares_2024b}.
These simulations have the same volume as the PM simulations used in this work, that is, a box side length of $328.125\, h^{-1}\mathrm{Mpc}$.
They evolve $512^3$ $N$-body particles in this volume using the code \texttt{ECOSMOG} \citep{Li_2012}, a modified-gravity version of \texttt{RAMSES} \citep{Teyssier_2002,Guillet_2011}.
The gravitational force is computed on a grid with adaptive mesh refinement (AMR), which achieves better spatial accuracy than the fixed grid of the PM simulations.
Dark matter haloes have been identified using a friends-of-friends (FoF) algorithm \citep{Davis_1985} with a linking length equal to $b=0.2$ times the mean interparticle distance, using the \texttt{pFoF} halo finder \citep{Roy_2014}.
From the halo positions, the overdensity fields are computed in the same way as for the matter fields described in \cref{subsec:train_samples}.
The main characteristics of these simulations are reported in \cref{tab:sims_summary}.
We refer the reader to \citet{Saez-Casares_2024b} for more details.

In order to build an emulator for the neural summary statistic, we use the \texttt{e-MANTIS} $w$CDM suite of simulations, with $80$ cosmological models covering the parameter ranges of \cref{table:cosmo_params}.
For each cosmological model, we average the summary statistic over the $64$ available independent realisations in order to reduce cosmic variance.
This significantly improves the accuracy of the emulator.
As in previous sections, we vary only $\Omega_{\rm m}$ and $\sigma_8$ during the cosmological inference and keep the other parameters fixed to their fiducial values.
We use the emulator validity range from \cref{table:cosmo_params} as flat priors for both of these parameters.

In order to compute the data covariance matrix, we use the set of $384$ independent realisations of the P18 reference cosmological model (see \cref{table:cosmo_params}).
We construct the fiducial data vector as the mean of all these realisations.
However, in the likelihood, we use the covariance of a single realisation, which has the same volume as in the previous sections.
Given that both the fiducial data vector and the emulator training data are averaged over a large number of realisations, we can consider them to be effectively noiseless.
Therefore, we once again ignore the $\boldsymbol{C}^{\rm sim}$ term in the total covariance as given by \cref{eq:covariance_total}.
This time, all the CNNs satisfy the Gaussian likelihood condition (see \cref{app:likelihood_validation}).
Therefore, we work with the full set of $20$ CNNs.

\begin{figure}
	\centering
	\includegraphics[width=\linewidth]{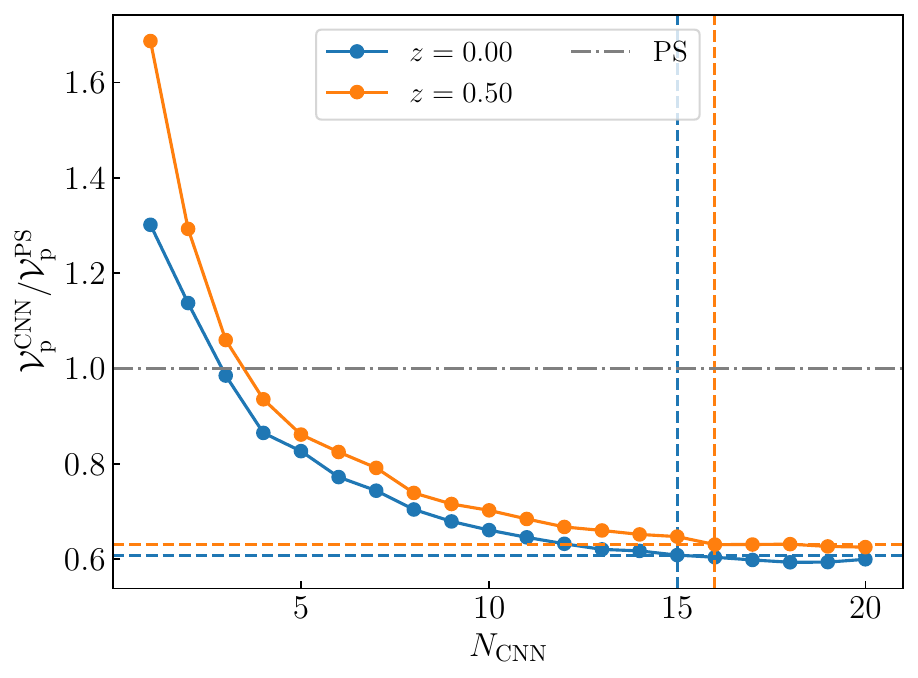}
	\caption{
	Posterior volume ratio as a function of the number of CNNs for different redshifts, using the greedy search approach.
	The number density is $n_{\rm h}=10^{-3}\,h^3\mathrm{Mpc}^{-3}$.
			The vertical and horizontal dashed lines mark the convergence point for each case.
			The dash-dotted line gives the power spectrum level, that is, a posterior volume ratio of one.
		}
	\label{fig:pvol_vs_n_cnn_emantis_z}
\end{figure}

We select haloes at a fixed number density.
More specifically, we select the $N_{\rm h}$ most massive haloes so as to reach the target number density $n_{\rm h}$.
This keeps the shot noise constant across all cosmologies, and we focus solely on the differences in clustering.
We consider three number density values $n_{\rm h}\in\left\{10^{-4}, 3\times10^{-4}, 10^{-3}\right\}\,h^3\mathrm{Mpc}^{-3}$.

\begin{figure*}
	\centering
	\begin{subfigure}{0.49\linewidth}
		\centering
		\includegraphics[width=\linewidth]{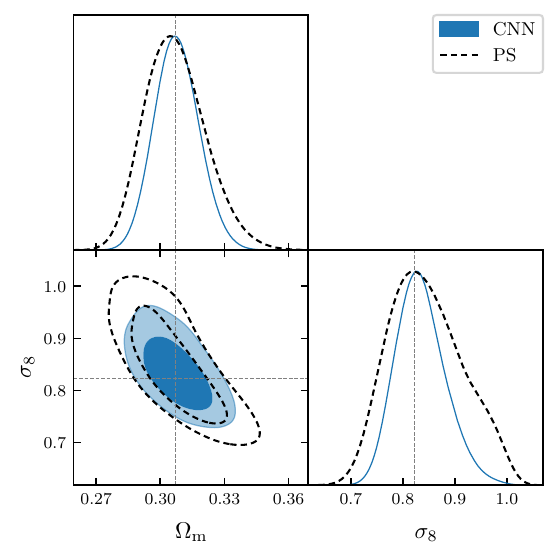}
	\end{subfigure}
	\begin{subfigure}{0.49\linewidth}
		\centering
		\includegraphics[width=\linewidth]{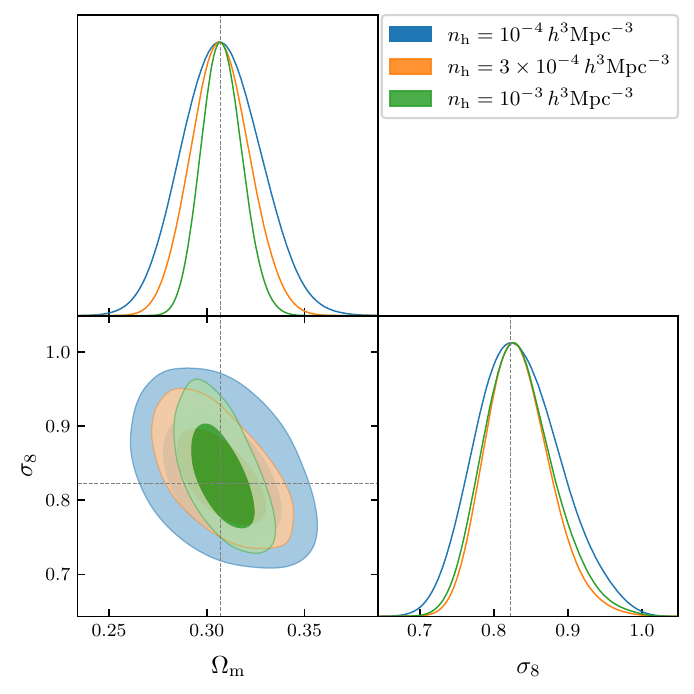}
	\end{subfigure}
	\caption{
	Two-dimensional posterior distributions of the cosmological parameters using $N_{\rm CNN}=20$ for the dark matter haloes.
	Left: comparison with power spectrum for the number density $n_{\rm h}=10^{-3}\,h^3\mathrm{Mpc}^{-3}$ at $z=0$.
				Right: neural summary analysis for different number density values at $z=0$.
		}
	\label{fig:posteriors_emantis}
\end{figure*}

In \cref{fig:pvol_vs_n_cnn_emantis_nd} we give the posterior volume ratio as a function of the number of networks used in the neural summary statistic, selected with the greedy search method, for different number densities at $z=0$.
The neural summary statistic achieves tighter cosmological constraints than the power spectrum.
The gain with respect to the power spectrum is larger for higher number densities.
Indeed, a larger number density corresponds to a denser sampling that includes lower-mass haloes and is expected to carry more beyond-two-point information.
Additionally, the densest sample has a lower shot noise, which gives access to smaller scales, where the neural summary is expected to provide the largest improvement over the power spectrum.
We also note that for higher number densities, the constraints converge with slightly fewer networks.

We repeat the same exercise, but considering the halo overdensity field at a higher redshift ($z=0.5$), where less beyond-two-point information is expected.
In \cref{fig:pvol_vs_n_cnn_emantis_z}, we show the results of the greedy search for the two redshift values $z \in \left\{0, 0.5\right\}$, considering the highest density bin $n_{\rm h}=10^{-3}\,h^3\mathrm{Mpc}^{-3}$.
The gain with respect to the power spectrum is slightly lower for the higher redshift bin, but the convergence in terms of the number of networks is similar.

In \cref{fig:posteriors_emantis}, we show the two-dimensional posteriors of the cosmological parameters for different types of analyses.
In the left panel, we compare the neural summary analysis to the power spectrum, in the case of the highest number density bin $n_{\rm h}=10^{-3}\,h^3\mathrm{Mpc}^{-3}$ at $z=0$.
In the right panel, we show the constraints from the neural summary analyses for different number density values at $z=0$.
\Cref{tab:summary_halo} summarises, for each cosmological parameter, the improvement in the standard deviation with respect to the power spectrum.
In all these cases, the full ensemble of $20$ networks has been used, so that the different analyses use the same number of networks.

\begin{table}[t]
	\centering
	\caption{
		Constraints from the neural summary relative to the power spectrum, for the $N$-body halo catalogues at different number densities and $z=0$.
	}
	\label{tab:summary_halo}
	\begin{tabular}{lccc}
		\hline
		\hline
		$n_{\rm h} \left(h^3\mathrm{Mpc}^{-3}\right)$ & $\sigma_{\Omega_{\rm m}} / \sigma_{\Omega_{\rm m}}^{P(k)}$ & $\sigma_{\sigma_8} / \sigma_{\sigma_8}^{P(k)}$ & $\mathcal{V}_{\rm p}/\mathcal{V}_{\rm p}^{P(k)}$ \\
		\hline
		$10^{-3}$                                     & $0.73$                                                     & $0.67$                                         & $0.60$                                           \\
		\hline
		$3\times10^{-4}$                              & $0.77$                                                     & $0.70$                                         & $0.68$                                           \\
		\hline
		$10^{-4}$                                     & $0.80$                                                     & $0.91$                                         & $0.83$                                           \\
		\hline
		\hline
	\end{tabular}
	\tablefoot{
		The ratios of one-dimensional standard deviations obtained with a neural summary analysis with respect to the power spectrum alone are given for each cosmological parameter, as well as the two-dimensional posterior volume ratio.
		The full ensemble of $20$ networks has been used in all cases.
	}
\end{table}

\section{Conclusions}
\label{sec:conclusions}

In this work, we have designed a neural summary statistic based on an ensemble of CNNs trained on a large number of affordable low-resolution PM simulations.
This statistic is able to extract information beyond the binned power spectrum when applied to higher-resolution simulations.

Each low-resolution simulation has a different combination of the five $\Lambda$CDM cosmological parameters ($\Omega_{\rm m}, \sigma_8, h, n_{\rm s}, \Omega_{\rm b}$) sampled from a Sobol sequence, and a different initial random seed.
From these, we computed the matter overdensity fields at $z=0$ on a regular grid with a Nyquist frequency of $k_{\rm Nyq}\simeq 0.6\,h\mathrm{Mpc}^{-1}$, and used them to train CNNs that compress the three-dimensional fields into a few coefficients.
Instead of considering a single network, we built an ensemble by varying the hyperparameters of the CNN architecture: the number of convolutional blocks, the number of convolutional layers per block, and the number of output channels in the first block.
Each network of the ensemble was trained independently from the others.

For cosmological inference, we applied the trained CNNs to the matter overdensity fields of high-resolution PM simulations and built a summary statistic by concatenating the outputs of a subset of networks selected from the ensemble.
The inference follows a standard Gaussian-likelihood pipeline, the only difference being that the field is compressed with our neural summary rather than with a standard statistic such as the power spectrum.
A Gaussian process emulator trained on high-resolution PM simulations provides the required theoretical predictions for the neural summary as a function of the cosmological parameters.
Its accuracy is accounted for through an emulator covariance matrix estimated with a $k$-fold cross-validation.
The data covariance matrix is computed from a dedicated suite of independent realisations of the fiducial P18 cosmology, one of which is used as the data vector.
We derived posteriors on $\Omega_{\rm m}$ and $\sigma_8$, keeping the other parameters fixed to their fiducial values.

We quantified the performance of the neural summary statistic by the posterior volume $\mathcal{V}_{\rm p}$ in the two-dimensional plane $\Omega_{\rm m}$--$\sigma_8$.
We compared three different methods for selecting the smallest subset of CNNs from the full ensemble that provides the best cosmological constraints: test set MSE ranking, individual posterior volume ranking, and greedy search.
We found that the greedy search provides the fastest convergence of the constraints with the number of networks used to build the neural summary.
In all cases, combining multiple CNNs greatly improves the cosmological constraints compared with a single network.
Combining the neural summary with the power spectrum yields only a minor further improvement, showing that the neural summary already captures most of the information contained in the power spectrum.

We then applied the neural summary to dark matter halo catalogues from high-resolution $N$-body simulations, at different redshifts and number densities.
Even though the CNNs were trained on low-resolution matter fields at a single redshift, the neural summary still extracts information beyond the binned power spectrum.
The improvement is larger at lower redshift and higher number density, where more beyond-two-point information is expected.
However, in this test the halo bias is not treated as a free parameter: its dependence on cosmology is captured by the emulator, which is built from the same halo selection, so that $\Omega_{\rm m}$ and $\sigma_8$ are the only free parameters for both the neural summary and the power spectrum analyses.
This test is therefore not representative of a realistic analysis of biased tracers, in which degeneracies between bias and cosmological parameters play an important role.

Overall, these results show that an ensemble of CNNs trained solely on fast but approximate low-resolution simulations can provide an informative summary statistic for cosmological inference from higher-fidelity data, including dark matter haloes, and that combining multiple networks is key to achieving this.
We stress that the CNN training samples differ substantially from the simulations used for inference: a low-resolution PM simulation requires about three orders of magnitude less computing time than a high-resolution one, and four orders of magnitude less than an $N$-body simulation with AMR.

In future work, we plan to extend this analysis to larger cosmological volumes, comparable to those of current galaxy surveys, and to more realistic synthetic galaxy catalogues, in order to assess the ability of the neural summary to break degeneracies between galaxy bias and cosmological parameters.
We also plan to explore the trade-off between the resolution and the number of training simulations, as well as different network architectures and strategies to generate the ensemble.

\begin{acknowledgements}
	This project has received funding from the European Union NextGeneration EU program – NRP Mission 4 Component 2 Investment 1.1 – MUR PRIN 2022 – Code 2022SKKYJN.
	Computational resources provided by INDACO Platform, which is a project of High Performance Computing at the University of MILAN (\url{http://www.unimi.it}).
	The authors acknowledge the computational resources provided by CINECA through the ISCRA initiative, using the Leonardo supercomputer.
	This project has benefited from computational time and data storage resources allocated by GENCI (Grand Équipement National de Calcul Intensif) at TGCC (Très Grand Centre de Calcul) for the projects 2021-A0110402287, 2022-A0130402287 and 2023-A0150402287 on the Joliot Curie supercomputer's ROME and SKYLAKE partitions.
  MC is partially supported by the 2025/26 ``Research and Education'' grant from Fondazione CRT.
  The OAVdA is managed by the Fondazione Cl\'ement Fillietroz, which is supported by the Regional Government of the Aosta Valley, the Town Municipality of Nus and the ``Unit\'e des Communes vald\^otaines Mont-\'Emilius''.
  We thank Pierre Zhang for useful discussions.
	This work has made use of the following numerical libraries on top of the ones already cited: \texttt{PyTorch}~\citep{pytorch}, \texttt{SciPy}~\citep{scipy}, \texttt{NumPy}~\citep{numpy}, \texttt{scikit-learn}~\citep{scikit-learn}, and \texttt{Matplotlib}~\citep{matplotlib}.
\end{acknowledgements}

\bibliographystyle{bibtex/aa}
\bibliography{bibtex/mybib}

\begin{appendix}

	\section{Gaussianity of the likelihood}
	\label{app:likelihood_validation}

	\begin{figure}
		\begin{center}
			\includegraphics[width=\linewidth]{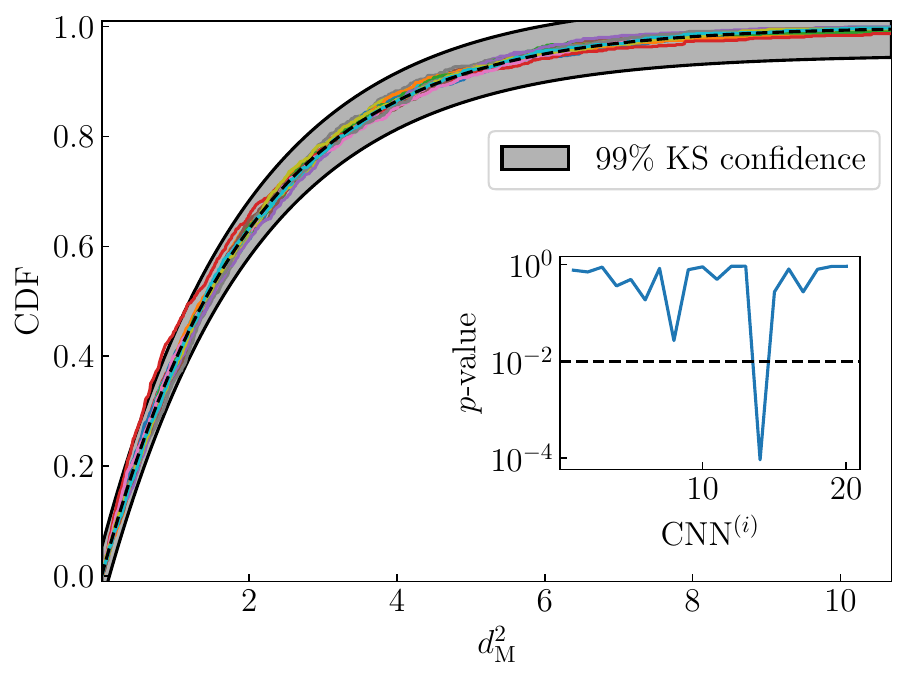}
		\end{center}
		\caption{
			Validation of the Gaussianity of the likelihood for the summary statistic built considering individual CNNs.
			The CDFs of the measured $d_{\rm M}^2$ values using the fields from the high-resolution PM simulations are shown, where each colour represents a different CNN.
			The theoretical chi-squared CDF with $\mathrm{df}=2$ degrees of freedom is given by the dashed line.
			The dark area marks the $99\%$ KS confidence region.
			The $p$-values associated with the KS test are reported in the inset for each CNN, where the dashed line marks the $p=0.01$ level.
		}
		\label{fig:likelihood_validation_single}
	\end{figure}

	Our cosmological analysis is based on the assumption that the summary statistic $\boldsymbol{t}$ is described by a multivariate Gaussian likelihood.
  To test this assumption, we compute the squared Mahalanobis distance \citep{Mahalanobis_1936} defined as
	\begin{equation}
		d_{\rm M}^2\left(\boldsymbol{t}_i\right) = \left(\boldsymbol{t}_i - \boldsymbol{\bar{t}}\right)^\top \cdot \boldsymbol{C}^{-1} \cdot \left(\boldsymbol{t}_i - \boldsymbol{\bar{t}}\right),
		\label{eq:chi2_likelihood_validation}
	\end{equation}
	where $\boldsymbol{t}_i$ is the summary statistic measured from the realisation $i$, $\boldsymbol{\bar{t}}$ is the mean computed over all realisations, and $\boldsymbol{C}^{-1}$ is the inverse of the covariance matrix estimated from all realisations.
	We compute $d_{\rm M}^2$ for each realisation from the fiducial cosmology set.
	If the statistic $\boldsymbol{t}$ follows a multivariate Gaussian distribution described by the covariance matrix $\boldsymbol{C}$, then the measured $d_{\rm M}^2$ should follow a chi-squared distribution with the number of degrees of freedom given by the length of $\boldsymbol{t}$.
	We use this property to validate the Gaussianity of the likelihood.

\begin{figure}
		\begin{center}
			\includegraphics[width=\linewidth]{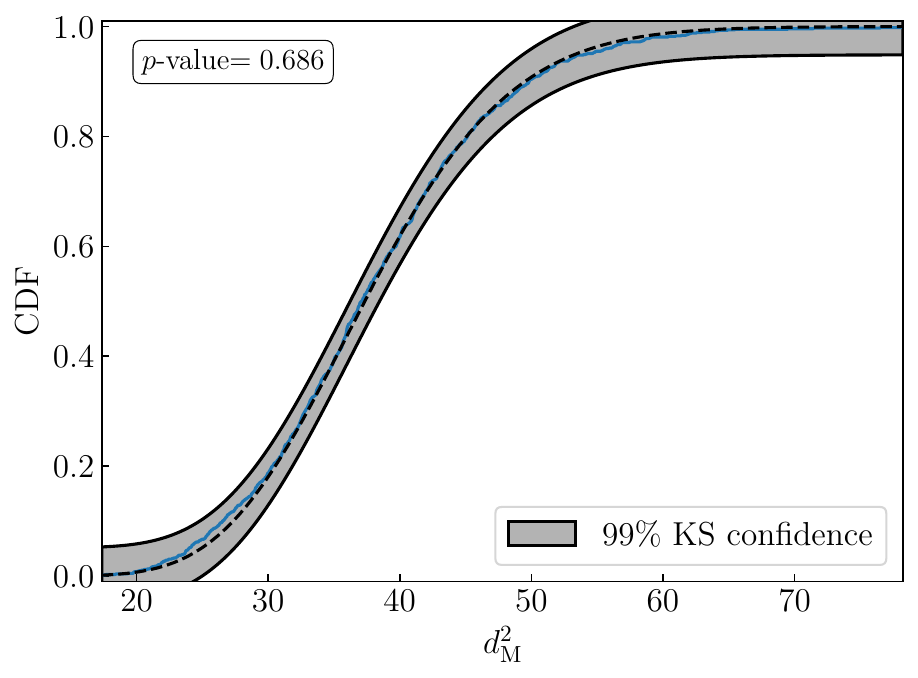}
		\end{center}
		\caption{
			Validation of the Gaussianity of the likelihood for the summary statistic built by concatenating the $19$ CNNs that satisfy the individual Gaussianity condition.
			The CDF of the measured $d_{\rm M}^2$ values using the fields from the high-resolution PM simulations is given by the solid line, while the theoretical chi-squared CDF with $\mathrm{df}=38$ degrees of freedom is given by the dashed line.
			The dark area marks the $99\%$ KS confidence region.
			The $p$-value associated with the KS test is reported.
		}
		\label{fig:likelihood_validation_all}
	\end{figure}

	Two distributions can be compared through their cumulative distribution functions (CDF).
	We show in \cref{fig:likelihood_validation_single} the CDF of measured $d_{\rm M}^2$ values along with the theoretical expectation of a chi-squared distribution.
	We focus in this case on the summary statistic $\boldsymbol{t}$ built from individual CNNs from the ensemble, when applied to the fields from the high-resolution PM simulations.
	Therefore, the number of degrees of freedom is $\mathrm{df}=2$.
	In this case, $1000$ realisations of the fiducial cosmology are available.
	We compute a Kolmogorov--Smirnov (KS) confidence region around the theoretical CDF, which accounts for the finite number of samples.
	For all the CNNs of the ensemble except one, the measured CDFs are within the $99\%$ confidence region.
	We note that the KS confidence region is not a pointwise confidence estimate but a global one, meaning that the full CDF should fit inside the confidence interval at a given confidence level.
	Alternatively, one can compute a $p$-value associated with the KS test, where the null hypothesis is that the measured samples are drawn from the reference distribution.
	The $p$-values for each CNN of the ensemble are reported in the inset of \cref{fig:likelihood_validation_single}, which shows more clearly that all CNNs except one pass the statistical test.
	Therefore, when using the high-resolution PM simulations for cosmological inference, we consider only $19$ CNNs out of the initial set of $20$.
	In \cref{fig:likelihood_validation_all} we show the results of the same kind of test but for the case of the summary statistic built by combining these $19$ CNNs.
	Overall, the Gaussianity of the likelihood is validated both for $19$ out of $20$ individual CNNs and for the combination of all $19$, within the statistical noise of the $1000$ realisations available for this test.

\begin{figure*}
		\centering
		\begin{subfigure}{0.49\linewidth}
			\centering
			\includegraphics[width=\linewidth]{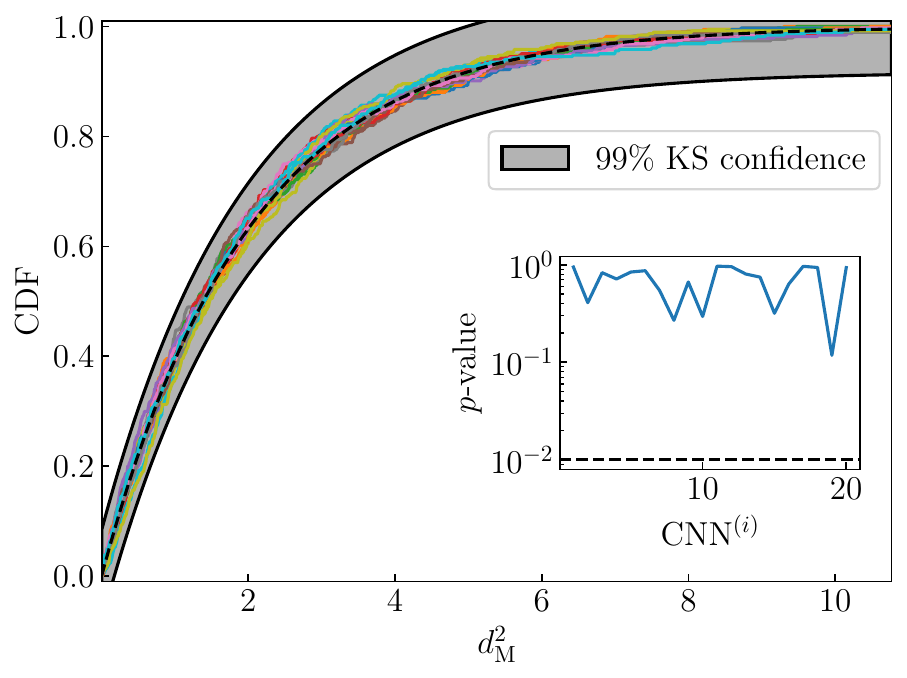}
		\end{subfigure}
		\begin{subfigure}{0.49\linewidth}
			\centering
			\includegraphics[width=\linewidth]{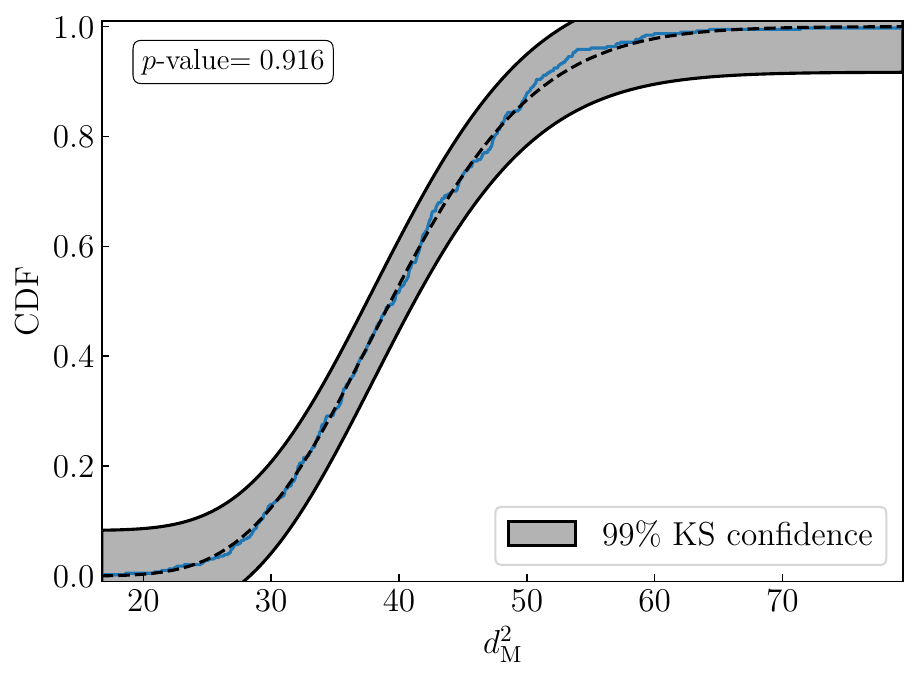}
		\end{subfigure}
		\caption{
		Same as \cref{fig:likelihood_validation_single} (\textit{left}) and \cref{fig:likelihood_validation_all} (\textit{right}) but for the $N$-body haloes from the \texttt{e-MANTIS} suite, for the number density bin $n_{\rm h}=10^{-3}\,h^3\mathrm{Mpc}^{-3}$ at $z=0$, in which case all networks are used ($\mathrm{df}=40$).
			}
		\label{fig:likelihood_validation_emantis}
	\end{figure*}

	Finally, we repeat the individual and combined CNN tests using the $N$-body haloes from the \texttt{e-MANTIS} simulation suite.
	We consider here a number density of $n_{\rm h}=10^{-3}\,h^3\mathrm{Mpc}^{-3}$ at $z=0$.
	The results are given in \cref{fig:likelihood_validation_emantis}.
	In this case, all the individual CNNs pass the likelihood Gaussianity test with large $p$-values.
	We therefore do not need to remove any CNN and can work with the whole ensemble.
	Moreover, the combination of all CNNs also passes the test.
	We find similar results for the other number density values and redshifts considered in this work.

	\section{Impact of the number of emulator training samples}
	\label{app:emu_nsims}

\begin{figure}
		\centering
		\includegraphics[width=\linewidth]{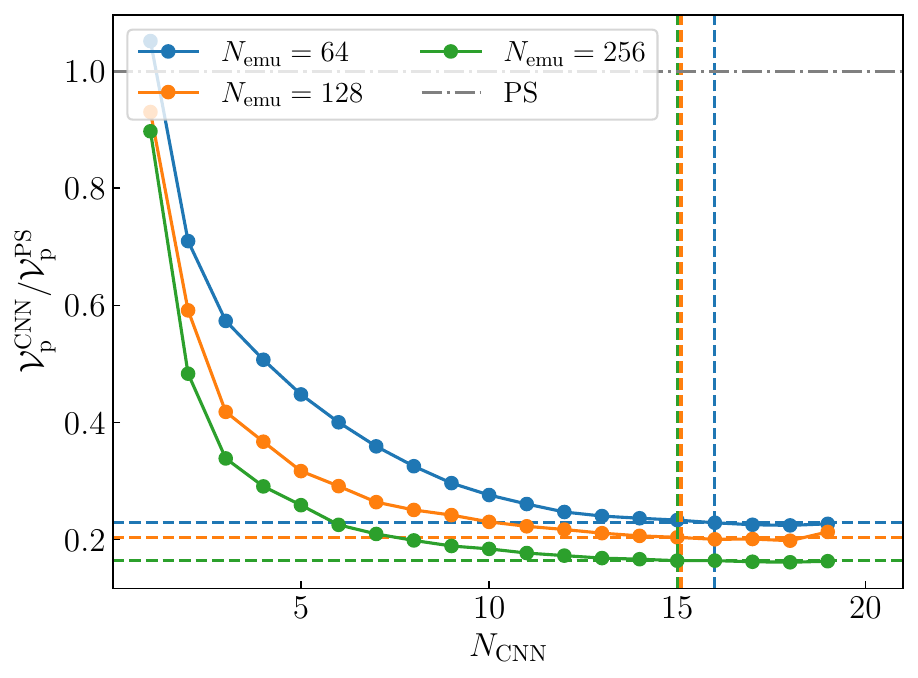}
		\caption{
			Posterior volume ratio as a function of the number of CNNs for different numbers of emulator training simulations.
			The vertical and horizontal dashed lines mark the convergence point for each case.
			The dash-dotted line gives the power spectrum level, that is, a posterior volume ratio of one.
		}
		\label{fig:pvol_vs_n_cnn_nemu}
	\end{figure}

	We investigate the impact of the number of training simulations for the emulator on the cosmological parameter constraints.
	For this, we repeat the greedy search procedure with different numbers of emulator training samples $N_{\rm emu}\in\{64,128,256\}$.
	The emulator covariance matrix is recomputed in each case with the $k$-fold cross-validation method, always using $16$ samples per fold.
	We present the results in \cref{fig:pvol_vs_n_cnn_nemu}.
	More specifically, we show the convergence of the posterior volume ratio with respect to the power spectrum case, as a function of the number of CNNs.
	For the power spectrum reference, we always consider the maximum number of emulator training simulations $N_{\rm emu}=256$.
	A larger number of emulator training simulations results in a smaller emulator error covariance matrix, which produces tighter cosmological constraints with the CNN-based analysis.
	The convergence of the constraints in terms of the number of networks is not significantly affected.

	\section{CNN training}
	\label{app:res_cnn_training}

	\begin{figure*}
		\begin{center}
			\includegraphics[width=\linewidth]{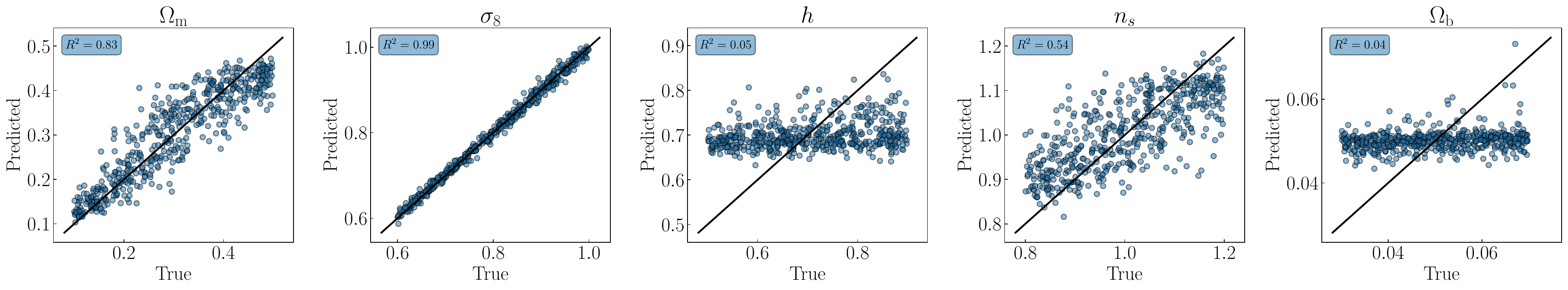}
		\end{center}
		\caption{
			Comparison between the CNN outputs and the true cosmological parameter values for the low-resolution test set simulations.
			The coefficient of determination $R^2$ is reported for each parameter.
			Here the case of the architecture with $\left\{N_{\rm conv\_block}=2, N_{\rm conv\_per\_block}=2, N_{\rm chs\_first}=5\right\}$ is shown as an example.
			Similar results are obtained for the other architectures.
		}
		\label{fig:cnn_pred_vs_true_test_set}
	\end{figure*}

	In this appendix we investigate the results of the CNN training procedure.
	For this, we study the output of the trained CNNs when applied to the low-resolution test set (see \cref{subsec:train_samples}), that is, to simulations consistent with the resolution of the training data.
	We recall that the CNN is trained to compress the input fields into the cosmological parameters (see \cref{subsec:loss_func}), which is equivalent to performing a regression of the cosmological parameters as a function of the input fields.
	In the context of building a neural summary statistic, this choice is the simplest approach, but other more complex and efficient alternatives exist \citep[see e.g.][]{Lanzieri_2025}.
	The test presented in this appendix is therefore specific to this choice.

	In order to quantify the quality of the regression, we compute for each target cosmological parameter the coefficient of determination defined as
	\begin{equation}
		\label{eq:r2score}
		R^2_j = 1 - \frac{\sum_{i=1}^{N_{\rm test}} (t^i_j- \theta^i_j)^2}{\sum_{i=1}^{N_{\rm test}} (\theta^i_j - \bar{\theta}_j)^2},
	\end{equation}
	where $\theta^i_j$ is the $j$-th cosmological parameter of sample $i$, $t^i_j$ is the corresponding CNN output, $N_{\rm test}$ is the number of samples in the test set, and the overbar denotes an average over the whole test set.
	A value of one indicates a perfect regression.
	We report these values in \cref{fig:cnn_pred_vs_true_test_set}, where we also show the predictions of the CNN for each target cosmological parameter as a function of the true values for each simulation of the test set.
	We focus here on the CNN architecture with $\left\{N_{\rm conv\_block}=2, N_{\rm conv\_per\_block}=2, N_{\rm chs\_first}=5\right\}$ as an example, but we obtain similar results for the other cases.

	The best measured cosmological parameters are $\Omega_{\rm m}$ and $\sigma_8$.
	Given the small cosmological volume of our simulation boxes, the CNN is not able to measure the cosmological parameters $h$, $n_{\rm s}$, or $\Omega_{\rm b}$.
	For this reason, we focus in this work on the parameters $\Omega_{\rm m}$ and $\sigma_8$.
	More specifically, when building the compressed summary statistic, we keep only the CNN outputs corresponding to these parameters and discard the rest, as explained in \cref{subsec:neural_summary_stat}.
	Additionally, during the cosmological inference, only these two parameters are varied and the rest are kept fixed to their fiducial values, as detailed in \cref{subsec:cosmo_posteriors}.

\end{appendix}

\end{document}